\documentclass[10pt,a4paper]{article}
\usepackage[latin1]{inputenc}
\usepackage{amsmath}
\usepackage{amsfonts}
\usepackage{amssymb}
\usepackage{graphics}
\usepackage{psfrag}
\usepackage{epsfig}
\usepackage{pstricks}
\usepackage{subfigure}

\newcommand{\ph}[1]{\phantom{#1}}
\newcommand{\w}{\wedge}
\newcommand{\ve}{\varepsilon}
\newcommand{\nn}{\nonumber}
\usepackage{amssymb}
\usepackage{color}
\usepackage{enumerate}
\def\be{\begin{equation}}
\def\ee{\end{equation}}
\def\bea{\begin{eqnarray}}
\def\eea{\end{eqnarray}}

\renewcommand{\texttt}{{}}

\begin{document}
\author{\\\\
Jo\~ao Magueijo $^1$\footnote{\texttt{j.magueijo@imperial.ac.uk}}\;, T.G. Zlosnik$^1$\footnote{\texttt{t.zlosnik@imperial.ac.uk}}\;, and T.W.B. Kibble$^1$\footnote{\texttt{t.kibble@imperial.ac.uk}}\\
{\small \it $(1)$ Imperial College Theoretical Physics, Blackett Laboratory, London, SW7 2AZ, United Kingdom}
}
\date{\today}

\title{Cosmology with a spin}

\maketitle

\begin{abstract} \noindent
Using the chiral representation for spinors we present a 
particularly transparent 
way to generate the most general spinor dynamics in a theory where gravity
is ruled by the Einstein-Cartan-Holst action. In such theories 
torsion need not vanish, but it can be re-interpreted as a 4-fermion
self-interaction within a torsion-free theory. The self-interaction 
may or may not break parity invariance, and may contribute positively
or negatively to the energy density, depending on the couplings 
considered. We then examine cosmological models ruled by a 
spinorial field within this theory. We find that  while there are cases for 
which no significant cosmological novelties emerge, the
self-interaction can also turn a mass potential into an upside-down Mexican hat
potential. Then, as a general rule, the model leads to cosmologies with
a bounce, for which there is a maximal energy density, and where the 
cosmic singularity has been removed. These solutions are stable, and range 
from the very simple to the very complex. 
\end{abstract}

\section{Introduction}

The greatest tragedy of XX century physics was by and large that gravity 
refused to partake in the successes of quantum field theory and
the gauge principle. This has led to numerous schemes purporting to supersede 
both classical relativity and standard quantization, but none of them was fully 
successful. A very basic question can be asked: if gravity is to be seen
as a gauge theory, which symmetry group is being gauged? It is possible to regard 
General Relativity as the symmetry broken phase of a gauge theory of groups 
for which the Lorentz group is a sub-group. Particular examples are
the Poincar\'e group \cite{Kibble:1961ba,Hehl:1994ue} and the de Sitter/anti-de Sitter
groups \cite{MacDowell:1977jt,Stelle:1979va,Randono:2010cq}.

A commonality of all these approaches is that they recover the so-called
`spin-connection' formulation of General Relativity, in which the gravitational
field is described by two independent ingredients. The first, referred to as the spin-connection, acts a gauge field for the Lorentz group whilst the second is a Lorentz valued spacetime one-form called the co-tetrad. It is from the
latter that the familiar metric tensor may be constructed. This formulation is elegant and desirable, but it does 
open up the doors to space-time torsion. In the presence of spinors one is naturally
led to build actions {\it directly} dependent on the spin-connection. These
produce a source term in the torsion equation of motion: and thus
torsion is forced upon gauge theories of gravity, whenever spinors are 
present. 

It turns out that, at least in the minimal theories with this
feature, the torsion is algebraically related to the spin density. Therefore 
it can explicitly be integrated out of the theory,
at least classically. It is found that the theory is equivalent 
to a torsion-free theory endowed with a 4-fermion self-interaction. 
However, even without considering more elaborate theories, with propagating
torsion for example, one has to face a number of different possibilities 
in the fermionic couplings. Terms which usually are boundary terms no longer
drop out of the equations whenever torsion is present. Therefore these
theories present a richness of possibilities, and the question naturally
arises as to how to constrain them. 

The first part of this paper (Sections~\ref{theory} and~\ref{eqsmot}) 
deals with the formal aspects of this matter. Availing ourselves of 
the chiral representation for spinors, we present a 
particularly transparent  way to generate the most general spinorial 
dynamics in a theory where gravity is ruled by the Einstein-Cartan-Holst 
action. We also work out explicitly the 4-fermion
self-interaction in the equivalent torsion-free theory. The self-interaction 
may or may not break parity invariance, and may contribute positively
or negatively to the energy density, depending on the couplings 
considered (see~\cite{Kibble:1961ba,Weyl:1950xa,Hehl:1971qi,Freidel:2005sn,Randono:2005up,Khriplovich:2005jh,Alexandrov:2008iy,Ellis:2011mz,Diakonov:2011fs} for related literature). 

In the second part of this paper (Sections~\ref{cosmoeqs},~\ref{ambid} and~\ref{parviol}) we examine cosmological models ruled by a  
spinor field within this theory. We find that the dynamics can be reduced
to a closed set of ODEs, representing the metric and some of the degrees of
freedom contained in the spinor bilinears (Section~\ref{cosmoeqs}). 
We then seek solutions to these equations. Solutions exhibiting parity-symmetry
(which we label ``ambidextrous'') are particularly simple to integrate. 
In the presence of torsion, for one sign for a given combination of 
the couplings, we find a bouncing Universe, driven by a torsion-induced
phantom phase  (Section~\ref{ambid}).
In these solutions there is a maximal energy density, and the 
cosmic singularity has been removed. These solutions are stable, and range 
from the very simple to the very complex, as we show for more general,
non-ambidextrous solutions in Section~\ref{parviol}.

In a concluding Section we list open issues to be addressed
in future work. We also include two Appendices where the notation used in 
this paper is thoroughly explained. 

\section{The theory}\label{theory}

In this paper we will look at the gravitational effect of a specific type of spinorial matter in cosmology. We first discuss our choice for the action describing the gravitational field. We shall use a first-order formalism for gravity where the gravitational field is described by the spacetime one-forms $e^{I}\equiv e^{I}_{\ph{I}\mu}dx^{\mu}$ and $\omega^{IJ}\equiv \omega^{IJ}_{\ph{IJ}\mu}dx^{\mu}$. The field $e^{I}$ is referred to as the co-tetrad and transforms homogeneously under local $SO(1,3)$ transformations whereas the field $\omega^{IJ}$, referred to as the spin-connection, transforms as a gauge field under similar transformations. We restrict ourselves to Lagrangians which are generally covariant, locally Lorentz invariant, and polynomial in our basic fields and their derivatives. Throughout this paper we shall rely heavily on the language of differential forms and we refer the reader to \cite{Westman:2012xk} for an introduction to these methods in gravitational theory. For compactness of notation we write the wedge product $a \w b$ of two differential forms simply as $ab$. The previous requirements on the gravitational Lagrangian restrict the number of possible terms considerably. Indeed it may be shown that up to boundary terms, the only `ingredients' that can be used are: co-tetrad $e^{I}$, the curvature of the spin-connection $R^{IJ} \equiv d\omega^{IJ} +\omega^{I}_{\ph{I}K} \omega^{KJ}$, 
and the $SO(1,3)$ invariant objects 
$\epsilon_{IJKL}$ and $\eta_{IJ}=\mathrm{diag}(-1,1,1,1)$ . 
Each of these quantities transform homogeneously under $SO(1,3)$ transformations and so one can combine these quantities to construct differential forms that are Lorentz scalars. We consider the following action:

\begin{eqnarray}\label{holst}
S_{G}[e^{I},\omega^{IJ}] &=& \kappa\int\left( \epsilon_{IJKL}+\frac{2}{\gamma} \eta_{IK}\eta_{JL}\right)e^{I} e^{J}R^{KL}
\end{eqnarray}
where $\kappa=1/(32\pi G)$.
The first term is the familiar Palatini action, whilst the second term is referred to as the Holst term, with $\gamma$ the Immirzi parameter. We do not include a cosmological constant term. The only further actions which are functionals only of $e^{I}$ and $R^{IJ}$ and polynomial in these fields are boundary terms quadratic in $R^{IJ}$ \cite{Rezende:2009sv}.


We now consider actions describing spinorial matter. If gravitation is essentially related to local Lorentz invariance then this matter will be described by Weyl spinors, i.e. vectors in the left and right handed representations of the group $SL(2,C)$ \cite{Palmer:2011bt}. We may now look to construct the most general action for $SL(2,C)$ spinors that produce the familiar spinor equations of motion in flat spacetime. We will use the label `($l$)' to denote quantities associated with the left handed representation and `($r$)' to denote quantities associated with the right handed representation. 
As in the gravitational case, we will look to construct actions which are polynomial in fields and locally Lorentz invariant. These actions should contain spacetime derivatives of the spinor fields so that they have dynamics. With this in mind, consider the following spacetime one-forms constructed from left handed Weyl spinors $\phi$ and right handed Weyl spinors $\chi$:

\begin{eqnarray}
K^{I(l)} &\equiv &\phi^{\dagger}\bar{\sigma}^{I}D^{(l)}\phi  \\
K^{I(r)} &\equiv & \chi^{\dagger}\sigma^{I}D^{(r)}\chi 
\end{eqnarray}
where 
\begin{eqnarray}
D^{(l)}\phi &\equiv &  d\phi-\frac{i}{2}\omega^{IJ}{\cal L}_{IJ}\phi+... \\
D^{(r)}\chi &\equiv & d\chi - \frac{i}{2}\omega^{IJ}{\cal R}_{IJ}\chi+...
\end{eqnarray}
The quantities ${\cal L}_{IJ}$ and ${\cal R}_{IJ}$ are the left and right handed generators of $SL(2,C)$ (see Appendix~\ref{conventions} for explicity expressions), whilst the ellipsis in each case denotes terms associated with coupling to different gauge fields. For instance, for spinor fields of the standard model of particle physics $D^{(l)}$ will contain the weak force gauge field whilst $D^{(r)}$ will not. Therefore $\phi$ and $\chi$ may have additional Yang-Mills indices, though we assume that the $\dagger$ operator itself has sufficient additional structure so that the kinetic terms are scalars under the relevant Yang-Mills transformations. The one-forms $K^{I(l)}$ and $K^{I(r)}$ transform as complex $SO(1,3)$ vectors and so in requiring real actions we may consider the following combinations:

\begin{eqnarray}
\label{s+}S_{(l)}[\phi,\omega^{IJ},e^{I}] &=& \int \epsilon_{JKLM}e^{J}e^{K} e^{L}\left[a_{(l)} \left(
K^{M(l)}+K^{M(l)*}\right)
 + ib_{(l)}\left(K^{M(l)}-K^{M(l)*}\right)\right]\\ 
\label{s-}S_{(r)}[\chi,\omega^{IJ},e^{I}] &=& \int \epsilon_{JKLM}e^{J} e^{K} e^{L}\left[a_{(r)} \left(
K^{M(r)}+K^{M(r)*}\right) 
+ ib_{(r)}\left(K^{M(r)}-K^{M(r)*}\right)\right]
\end{eqnarray}
where the $a_{l,r}$ and $b_{l,r}$ are real constants. We note by inspection that the following relations hold:

\begin{eqnarray}
K^{I(l)}+K^{I(l)*} &=& D(\phi^{\dagger}\bar{\sigma}^{I}\phi) \\
 K^{I(r)}+K^{I(r)*} &=& D(\chi^{\dagger}\sigma^{I}\chi)
\end{eqnarray}
Therefore the $a_{(l)}$ and $a_{(r)}$ terms of the action may be written collectively as follows:

\begin{eqnarray}
\int\epsilon_{IJKL}e^{I}e^{J} e^{K} D\left( a_{(l)}\phi^{\dagger}\bar{\sigma}^{L}\phi+
a_{(r)}\chi^{\dagger}\sigma^{L}\chi\right) 
&=& 3\int \left( a_{(l)}\phi^{\dagger}\bar{\sigma}^{L}\phi+
a_{(r)}\chi^{\dagger}\sigma^{L}\chi\right)\epsilon_{IJKL} T^{I} e^{J} e^{K} \nn \\
&& -\int d\left[\epsilon_{IJKL}\left( a_{(l)}\phi^{\dagger}\bar{\sigma}^{L}\phi+
a_{(r)}\chi^{\dagger}\sigma^{L}\chi\right)e^{I} e^{J} e^{K}\right] \label{act1}
\end{eqnarray}
where

\begin{eqnarray}
T^{I} &\equiv & De^{I} = de^{I}+\omega^{I}_{\ph{J}J} e^{J}
\end{eqnarray}
The two-form $T^{I}$ is referred to as the spacetime torsion and is assumed to be vanishing in the conventional second-order metric formulation of gravity \cite{Wald:1984rg}. We see from (\ref{act1}) that the $a_{(l)}$ and $a_{(r)}$ only contribute to the equations of motion (i.e. are not described only by a boundary term) when $T^{I}(x^{\mu})$ is non-vanishing. We may additionally consider an action $S_{U}[\phi,\chi,e^{I}]$ describing  `potential' terms built from combinations of $\phi$ and $\chi$:

\begin{eqnarray}
S_{U} &= & -\frac{1}{4!}\int U(\phi,\chi) \epsilon_{IJKL}e^{I} e^{J} e^{K} e^{L}
\end{eqnarray}
Note that the inclusion of factors such as $4!$ in the above equation and $3!$ later on are to ensure neatness when actions are written in standard form (see Appendix~\ref{standard}).
Our total spinor action then is $S_{\Psi}\equiv S_{(l)}+S_{(r)}+S_{U}$, i.e.

\begin{eqnarray}
\nn S_{\Psi}[e^{I},\omega^{IJ},\phi,\chi] &=& \int \epsilon_{IJKL}e^{J} e^{K}  \biggl\{i e^{I}  \left[b_{(l)} \left(
K^{L(l)}-K^{L(l)*}\right) + b_{(r)}\left(K^{L(r)}-K^{L(r)*}\right)\right]\\
&&+3 T^{I}\left( a_{(l)}\phi^{\dagger}\bar{\sigma}^{L}\phi+
a_{(r)}\chi^{\dagger}\sigma^{L}\chi\right) - \frac{1}{4!}U e^{I} e^{L}\biggr\} 
\end{eqnarray}
Additional notational simplification is possible if we describe spinors in terms of Dirac spinors. Following the field redefinition $\chi\rightarrow \sqrt{b_{(l)}/b_{(r)}}\chi$ we define the Dirac spinor $\Psi= (\phi_{\alpha},\chi^{\alpha'})$ and so the action $S_{\Psi}$ becomes:

\begin{eqnarray}
S_{\Psi}[e^{I},\omega^{IJ},\Psi] &=& \frac{1}{3!}\int \epsilon_{IJKL}e^{J} e^{K}  \left(i e^{I}  \left(\frac{1}{2}\left(
\bar{\Psi}\gamma^{L}D\Psi - D\bar{\Psi}\gamma^{L}\Psi\right)\right)+ \frac{3}{2}T^{I}\left( \alpha V^{L}+\beta A^{L}\right)\right)\nn\\
&& - \frac{1}{4!}\int U\epsilon_{IJKL}e^{I} e^{J} e^{K} e^{L}+S_{int}(\chi,\phi,B_{\mu},..)
\end{eqnarray}
where $\bar{\Psi}\equiv \Psi^{\dagger}\gamma^{0}$ and $U$ is specifically considered a  function formed from scalars formed from Dirac spinors and the basic objects of $spin(1,3)\simeq SL(2,C)$ e.g. $\bar{\Psi}\Psi$ or $\bar{\Psi}\gamma^{5}\Psi$. The covariant derivative is defined as $D\Psi\equiv d\Psi- \frac{i}{2}\omega^{IJ}{\cal J}_{IJ}\Psi$ where ${\cal J}_{IJ}$ are the generators of $spin(1,3)$ (see Appendix~\ref{conventions}). The action $S_{int}$ contains  interaction terms between $\phi$, $\chi$, and Yang-Mills gauge fields $B_{\mu}$.
Additionally we have defined the following quantities: 

\begin{eqnarray*}
V^{L} &\equiv & \bar{\Psi}\gamma^{L}\Psi=\phi^{\dagger}\bar{\sigma}^{L}\phi+\chi^{\dagger}\sigma^{L}\chi 
\quad\quad A^{L} \equiv  \bar{\Psi}\gamma^{5}\gamma^{L}\Psi=\phi^{\dagger}\bar{\sigma}^{L}\phi-\chi^{\dagger}\sigma^{L}\chi \\
b_{(l)} &\equiv& \frac{1}{12}
\quad\quad\alpha \equiv\frac{1}{12}\left(a_{(l)}+a_{(r)}\frac{b_{(l)}}{b_{(r)}}\right) 
\quad\quad \beta \equiv\frac{1}{12}\left(a_{(l)}-a_{(r)}\frac{b_{(l)}}{b_{(r)}}\right)
\end{eqnarray*}
The vectors $V^{L}$ and $A^{L}$ are referred to, respectively, as the vector and axial current density of the spinor field $\Psi$.
These quantities will be of particular importance. One may additionally consider terms coupling spinor invariants to curvature and torsion, for instance $\bar{\Psi}\Psi T^{I} T_{I}$, but 
we will not consider them in this present work. Our combined action will therefore be $S \equiv S_{G}+S_{\Psi}$. For ease of comparison with similar actions considered in the literature, we present the form of this action in `standard' tensor notation in Appendix~\ref{standard}.

\section{Equations of motion and first implications}\label{eqsmot}

We now derive the equations of motion. As usual these will be defined by the requirement of stationarity of the action $S$ under small variations of the dynamical fields. Varying with respect to the spin-connection $\omega^{IJ}$ we have that:

\begin{eqnarray}
\frac{\delta S}{\delta \omega^{KL}} = 0&=& -2 \kappa\left( \epsilon_{IJKL}+\frac{2}{\gamma}\eta_{I[K}\eta_{L]J}\right)T^{I} e^{J}+\frac{1}{4!}\epsilon_{IMNP}e^{I} e^{M}  e^{N}\varepsilon^{DP}_{\phantom{DP}KL}A_{D} \nonumber \\
&& -\frac{1}{4}\epsilon_{[K|MNQ}e^{M} e^{N} e_{|L]}\left(\alpha V^{Q}+\beta A^{Q}\right) \label{toreq}
\end{eqnarray}
Therefore we see that the torsion $T^{I}$ is sourced by axial \emph{and} vector currents $A^{L}$ and $V^{L}$.
For computational convenience it will be useful to decompose the spin-connection as follows $\omega^{IJ}=\tilde{\omega}^{IJ}+C^{IJ}$. The quantity $\tilde{\omega}^{IJ}$ is defined to be the solution to the spin connection 
when $V^{L}=A^{L}=0$, i.e. it is a solution to the equation $T^{I}=0$. Therefore $\tilde{\omega}^{IJ}$ depends only upon $e^{I}$ and its partial derivatives. By implication the torsion may be expressed in terms of the contorsion one-form $C^{IJ}=C^{IJ}_{\ph{IJ}\mu}dx^{\mu}$  as follows: $T^{I} = C^{IJ} e_{J}$. Furthermore we define the contorsion scalar $C^{IJM}e_{M\mu}dx^{\mu} \equiv C^{IJ}_{\ph{IJ}\mu}dx^{\mu}$; after calculation it may be seen that (\ref{toreq}) implies the following solution for this quantity:

\begin{eqnarray}
C_{TLK} &=& \frac{\gamma^{2}}{8\kappa\left(\gamma^{2}+1\right)}\left[\ve^{D}_{\ph{D}TLK}\frac{1}{2}\left(A_{D}+\frac{1}{\gamma}\left(\alpha V_{D}+\beta A_{D}\right)\right)\right.  \nn\\
        &&  \left.-\frac{1}{\gamma} A_{[L}\eta_{T]K}+\alpha V_{[L}\eta_{T]K}+\beta A_{[L}\eta_{T]K}\right]
\end{eqnarray} 
In summary then we have solved for $\omega^{IJ}$ in terms of $\tilde{\omega}^{IJ}$ and $C^{IJ}$, which themselves may be expressed entirely in terms of $e_{I}$, $V^{I}$, $A^{I}$ and their derivatives. In this sense we have eliminated torsion from the theory as the remaining equations of motion may be expressed entirely in terms of variables familiar from the second-order formalism.
We now find the Einstein equations which follow from varying the action with respect to the co-tetrad $e^{I}$:

\begin{eqnarray}
\nn \frac{\delta S}{\delta e^{I}}= 0 &=& 2\kappa\epsilon_{JIKL}e^{J} \tilde{R}^{KL}-\frac{i}{4}\epsilon_{IJKL}e^{J} e^{K}   \left(
\bar{\Psi}\gamma^{L}\tilde{D}\Psi - \tilde{D}\bar{\Psi}\gamma^{L}\Psi\right)\\
\nn&& -\frac{U}{6}\epsilon_{MJIL}e^{J} e^{M}  e^{L}+2\kappa\left( \epsilon_{JIKL}+\frac{2}{\gamma}\eta_{JK}\eta_{IL}\right) C^{K}_{\ph{K}MP}C^{ML}_{\ph{ML}Q}e^{J} e^{P} e^{Q}\\
\nn && +\frac{1}{2}\epsilon_{IKNL}\left(\frac{1}{4} \ve^{EL}_{\ph{EL}AB}A_{E}+\frac{1}{2}\delta^{L}_{A}(\alpha V_{B}+\beta A_{B})\right)C^{AB}_{\ph{AB}P}e^{N} e^{K} e^{P} 
\end{eqnarray}
%
Objects with a `$\tilde{\ph{a}}$' above them denote quantities constructed from the `zero-torsion' spin connection, obtained
by replacing $\omega^{IJ}$ with $\tilde{\omega}^{IJ}$ wherever they occur within the object. Finally we write down the spinor equation of motion which comes from considering variations with respect to $\bar{\Psi}$:

\begin{eqnarray}
\nn \frac{\delta S}{\delta \bar{\Psi}} =0 &=& 4i\epsilon_{IJKL}e^{J} e^{K} e^{I}\gamma^{L} D\Psi+6\alpha\epsilon_{IJKL}e^{J} e^{K} T^{I}\gamma^{L}\Psi \\
&& + 6\left(1+\beta\right)\epsilon_{IJKL}e^{J} e^{K} T^{I}\gamma^{5}\gamma^{L}\Psi-\frac{\delta U}{\delta \bar{\Psi}}\epsilon_{IJKL}e^{J} e^{K} e^{I} e^{L}
\end{eqnarray}

\subsection{The four-fermion interaction}

It is instructive to write the Einstein and Dirac equations in standard tensor notation. By calculation we find that:

\begin{eqnarray}
\nn 4\kappa \tilde{G}_{\mu\nu} &=& -\frac{i}{2}e_{L\mu}\left(\bar{\Psi}\gamma^{L}\tilde{D}_{\nu}\Psi-\tilde{D}_{\nu}\bar{\Psi}\gamma^{L}\Psi\right) + \frac{i}{2}e^{\sigma}_{L}\left(\bar{\Psi}\gamma^{L}\tilde{D}_{\sigma}\Psi-\tilde{D}_{\sigma}\bar{\Psi}\gamma^{L}\Psi\right)g_{\mu\nu} \\
&& - W g_{\mu\nu}\\
\label{dirnor} i\gamma^{L}e^{\mu}_{L}\tilde{D}_{\mu}\Psi &=& \frac{\delta{W}}{\delta\bar\Psi}
\end{eqnarray}
where $W$ is the `effective potential', incorporating the effects of the non-vanishing contorsion form:

\begin{eqnarray}\label{4ferm}
W \equiv  U +\frac{3\pi G\gamma^{2}}{2\left(1+\gamma^{2}\right)}\left[\left(1-\beta^{2}+\frac{2}{\gamma}\beta\right)A_{I}A^{I}-\alpha^{2}V_{I}V^{I}-2\alpha\left(\beta-\frac{1}{\gamma}\right)A_{I}V^{I}\right] 
\end{eqnarray}
(Here we have used the fact that $\kappa=1/32\pi G$.) For simplicity we restrict ourselves to the case where the potential $U$ depends only on $\bar{\Psi}\Psi$. We may use the Dirac equation to simplify the Einstein equations somewhat, recasting them in the following form:

\begin{eqnarray}
\label{einnor} \frac{1}{8\pi G}\tilde{G}_{\mu\nu} &=& -\frac{i}{2}e_{L\mu}\left(\bar{\Psi}\gamma^{L}\tilde{D}_{\nu}\Psi-\tilde{D}_{\nu}\bar{\Psi}\gamma^{L}\Psi\right)+\left(W+\frac{\partial U}{\partial (\bar\Psi\Psi)}\bar{\Psi}\Psi-2U\right)g_{\mu\nu} 
\end{eqnarray}
Equations (\ref{dirnor}) and (\ref{einnor}) are the classical equations of motion for a Dirac spinor field non-minimally coupled to gravity.

Equation~(\ref{4ferm}) is the central result in the first part of our 
paper (theoretical set up). It shows that the torsion effects predicted
by our theory can be recast in the form of 4-fermion self-interactions.
This is not new (see~\cite{Weyl:1950xa,Kibble:1961ba,Hehl:1971qi} 
for instance), but we have applied this idea to a more general framework.
It is useful to consider the various limits
of the theory. Minimal coupling for the spinor is obtained by removing 
terms in (\ref{s+}) and (\ref{s-}) which are pure boundary terms when 
the torsion vanishes. This amounts to setting $a_{(l)}=a_{(r)}=0$, i.e. 
$\alpha=\beta=0$. More generally, when $\gamma=0$, the self-interaction 
vanishes, and we recover the torsion-free, second order formulation. 
Indeed setting $\gamma=0$ 
in (\ref{holst}) is equivalent to setting the term multiplying $1/\gamma$
to zero, i.e. setting the torsion to zero; this is the theory underlying the
cosmological models studied in~\cite{ArmendarizPicon:2003qk}.
In contrast, when $\gamma\rightarrow \infty$ we recover 
Einstein-Cartan-Sciama-Kibble theory for which the cosmological effect of spinorial matter has been considered in some detail \cite{Bauerle:1983ai,Dolan:2009ni,Watanabe:2009nc}. However vector-vector and a parity
violating axial-vector interactions do appear for more general couplings,
as we have shown here.

In between these extreme cases we find a class of theories parametrized
by the Immirzi parameter and the non-minimal coupling constants 
$\alpha$ and $\beta$. The form of $W$ for the case where $\beta=0$ has been
worked out \cite{Freidel:2005sn,Randono:2005up,Khriplovich:2005jh}, 
and our general
result falls within the results in~\cite{Alexandrov:2008iy,Diakonov:2011fs}. The exotic case $\gamma=\pm i$ sees the interaction
diverge. This corresponds to setting the (anti)self-dual
current to zero in the minimally coupled case, and more generally
what's inside the bracket in Eq.~(\ref{4ferm}). However, our insistance upon real actions restricts us to real values for $\gamma$.

\subsection{An application: a Classical Spinor Field}\label{signals}

It is expected that the equations of motion (\ref{dirnor}) and (\ref{einnor}) should ultimately be regarded
as operator equations for the spinorial and gravitational quantum fields. We shall simply assume that 
the classical gravitational field that we observe is sourced in (\ref{einnor}) by expectation values of
spinor invariants such as $\bar{\Psi}\Psi$ and $A_{I}A^{I}$. It was noted in \cite{Dolan:2009ni} that in
the first-order formulation of gravity there is an ambiguity in taking expectation values, wherein it is 
arguably more natural to take expectation values in the equation of motion $\delta S/\delta \omega^{IJ}=0$. Upon
solving for the (classical) contorsion, this would yield contributions proportional to $\left<A_{I}\right>\left<A^{I}\right>$. 
If, however, one had started from the second-order formalism with four-fermion interaction, it would be expected that contributions in the Einstein equations would be of the form $\left<A_{I}A^{I}\right>$. This is an important issue as 
$\left<A_{I}\right>\left<A^{I}\right>$ and $\left<A_{I}A^{I}\right>$ will typically not be identical \cite{Hehl:1976kj}.
We note that our ability to cast our model as a second order theory has relied on being able to solve for the contorsion form algebraically; for modest modifications to gravity this is no longer possible (see for instance \cite{Alexander:2008wi}) and so
one may doubt the primacy of the second-order formulation of gravity and its implications for the gravitational effect of fermions.

We circumvent this possible ambiguity by restricting ourselves to spinors called \emph{classical spinors}. These 
are defined to be quantum fields described by a spinorial operator $\Psi=(\phi,\chi)$ and assumed to be in a state where the expectation value $\left<f(\Psi)\right>\approx f(\left<\Psi\right>)$. If this is the case then the above ambiguity in the averaging of $A_{I}A^{I}$ disappears. Henceforth we will confine ourselves to situations where the approximation $\approx$ is sufficiently good to be regarded as equality, and the field $\left<\Psi\right> \equiv \Psi_\mathrm{cl}$ 
will be referred to as a classical spinor. It should be noted that familiar fields such as those describing quarks and leptons are not classical in this sense
on cosmological scales, therefore when we consider a spinor field that is independent from the fields
of the standard model. If we consider explicit components (in the representation of the Dirac matrices given in Appendix A), 
$\Psi_\mathrm{cl} =(a,b,c,d)$, where $a,b,c,d$ are assumed to be complex numbers, we have that:

\begin{eqnarray}
\left<A_{I}A^{I}\right> &=& 4(a^{*}c+b^{*}d)(a^{*}c+b^{*}d)^{*} \\
\left<V_{I}V^{I}\right> &=& -4(a^{*}c+b^{*}d)(a^{*}c+b^{*}d)^{*} \\
\left<V_{I}A^{I}\right> &=& 0\\
\left<\bar{\Psi}\Psi\right> &=& (a^{*}c+b^{*}d)+(a^{*}c+b^{*}d)^{*}\\
\left<\bar{\Psi}\gamma^{5}\Psi\right> &=& (a^{*}c+b^{*}d)-(a^{*}c+b^{*}d)^{*}
\end{eqnarray}
Therefore, for the classical spinor, a non-vanishing $\left<A_{I}A^{I}\right>$ is always spacelike, a non-vanishing $\left<V_{I}V^{I}\right>$ is always timelike, and they are of equal magnitude. All of the above quantities depend upon a single complex number:

\begin{eqnarray}
a^{*}c+b^{*}d\equiv \frac{1}{2}\left(E+Bi\right)
\end{eqnarray}
hence we have:

\begin{eqnarray}
\left<A_{I}A^{I}\right> = -\left<V_{I}V^{I}\right> &=& (E^{2}+B^{2}) \\
\left<\bar{\Psi}\Psi\right> &=& E \\
\left<\bar{\Psi}\gamma^{5}\Psi\right> &=& iB
\end{eqnarray}
We have explicitly derived these identities in order to motivate the introduction of variables $E$ and $B$, but they could have been obtained more directly with knowledge of the Pauli-Fierz relation:
\be({\bar \Psi} Q \gamma_I \Psi) ({\bar \Psi} Q \gamma^I \Psi) =-({\bar \Psi} Q \Psi)({\bar \Psi} Q \Psi) + ({\bar \Psi} Q \gamma_5 \Psi)( {\bar \Psi} Q \gamma_5 \Psi)\; ,
\ee 
where $Q\in\{1,\gamma_5\}$. Given the assumption of a classical spinor, the form of the function $W$ simplifies considerably:

\begin{eqnarray}
W &= &  U(E)+ \xi \left(E^{2}+B^{2}\right)
\end{eqnarray}
where
\begin{eqnarray}\label{xi}
\xi =\frac{3\pi G\gamma^{2}}{2(\gamma^{2}+1)} \left(1+\alpha^{2}+\frac{2}{\gamma}\beta-\beta^{2}\right)\; .
\end{eqnarray}
We stress that within the set of couplings considered $\xi$ can be
positive or negative. Recalling that $A^I$ must be space-like, this
means that the contribution to the overall energy due to torsion may be
positive or negative in our model, a fact that will have far reaching
consequences in this paper.

\section{Cosmological equations}\label{cosmoeqs}
We would like to set up a model based on the FRW metric and a classical spinor
field, sourcing torsion. It is not immediately obvious that this is 
possible. Consider the axial current
$A^I$ for a generic classical spinor. Since this is space-like, there isn't 
a frame where its spatial components vanish, and therefore any spinor field
picks up a preferred direction. However, this does not imply that the 
metric has to be anisotropic. In fact, as Isham and Nelson
showed \cite{Isham:1974ci}, the metric may still be the FRW metric even if 
the spinor is anisotropic, as long as $K= 0$, i.e. it is the spatially flat
FRW metric. Otherwise it is impossible to satisfy the Einstein equations, 
precisely because the $A^{i}$ cannot be made to vanish. We shall therefore 
assume $K=0$ throughout this paper. In a sense this is a solution to the 
flatness problem: the existence of a spinor in a Friedmann universe 
would preclude the existence of spatial curvature. 

The cosmological consequences of a classical spinor have variously been considered
in the literature. In \cite{ArmendarizPicon:2003qk} the cosmological effect 
of a classical spinor with potential $U(\bar{\Psi}\Psi)$ and with $\gamma=0$ 
(i.e. vanishing torsion) up to small perturbations around a spatially flat FRW 
universe was considered. This analysis has subsequently been extended to the Einstein-Cartan
minimal coupling case $\gamma=\infty$, $\alpha=\beta=0$ \cite{Watanabe:2009nc}. 
Further to this, non-minimal coupling $\alpha\neq0,\beta=0$ has been considered
at the level of the cosmological background in \cite{deBerredoPeixoto:2012xd} though we note
that our sign for the four-fermion interaction in $W$ does not agree.

We make the following choice for our co-tetrads: $e^{0}=dt$, 
$e^{i}=a(t)dx^{i}$ where indices $i,j,k,..$ go from 1 to 3. We first 
obtain the non-vanishing components of the field $\tilde{\omega}^{IJ}$, 
the torsion-free spin connection, by solving 
$de^{I}+\tilde{\omega}^{I}_{\phantom{I}J} e^{J}=0$.
By inspection we have $de^{I}= Ha\delta^{I}_{i}dt dx^{i}$ where 
$H\equiv \frac{\dot{a}}{a}$ and so we have
%
$\tilde{\omega}^{i}_{\phantom{i}0}=He^{i}$. 
Given our ansatz for the spinor field and geometry, it may be shown 
that the Dirac equation takes the following form:
\begin{eqnarray}
\gamma^{0}\left(\dot{\Psi}+\frac{3}{2}H\Psi\right) &=& - i\frac{\delta W}{\delta \bar{\Psi}}
\end{eqnarray}
and taking the Hermitian conjugate of the above equation we get
\begin{eqnarray}
\left(\dot{\Psi}^{\dagger}+\frac{3}{2}H\Psi^{\dagger}\right)\gamma^{0} &=&  i\left(\frac{\delta W}{\delta \bar{\Psi}}\right)^{\dagger}
\end{eqnarray}
The simple algebraic 
facts we have just presented are enough to derive a closed set
of ordinary differential equations for the spinor and gravitational fields. 
We choose to express the spinor dynamics in terms of its quadratic
invariants, since these are the observables of the theory (rather than
the spinor field itself). Indeed it turns out that 
a complete closed set of equations for our system, assuming the FRW metric 
with $K=0$, is formed by:
\begin{eqnarray}
\label{dote}\dot{E}+3HE &=& 4\xi BA^{0} \\
\label{dotb}\dot{B}+3HB &=& -4\xi  EA^{0}- 2 U'(E) A^{0} \\
\label{dota}\dot A^{0}+3HA^{0} &=& 2 U'(E)B \\
\label{fried} H^{2} &=& \frac{8\pi G}{3}\left(U +\xi \left(E^{2}+B^{2}\right)\right) 
\equiv \frac{8\pi G}{3}\rho
\end{eqnarray}
%
where the prime denotes differentiation with respect to $E$. This is a minimal set specifying the dynamics, and will form the basis for a numerical study in Sections~\ref{ambid} (where some of equations become trivial) and \ref{parviol}. 
Other equations could be added to this set. For example:
\be
\dot V^{0}+ 3HV^{0} = 0
\ee
but this drops out from the gravitational 
dynamics altogether. Likewise one could
write more ODEs ruling the dynamics of the remaining variables associated 
with the bilinears, but they also drop out from the cosmological dynamics. 
For completeness, the second Friedmann equation is given by
\be
-2\frac{\ddot{a}}{a}-H^{2} 
=8\pi G \left[U'E-U+\xi \left(E^{2}+B^{2}\right)\right] \equiv 8\pi G p\; .
\ee
This equation can be derived from (\ref{fried})
and conservation equation 
\be\label{cons}
\dot \rho + 3H(p+\rho)=0
\ee 
(written in terms of $E$ and $B$),
which in turn follows from the (\ref{dote}), (\ref{dotb}) and (\ref{dota}).
This is, however, not true at turn-around points (where $H=0$), because
the conservation equation becomes degenerate ($0=0$). The second Friedmann
equation is then needed, and failure to take this fact into account might
lead one to mistake a bounce for a static or loitering Universe. Finally we note that equations (\ref{dote}), (\ref{dotb}) 
and (\ref{dota}) provide a first integral:
\be\label{firstint}
a^6 [E^2 + B^2 + (A^0)^2)]= M^2
\ee
This equation was first noted in \cite{deBerredoPeixoto:2012xd}.

\section{Ambidextrous solutions}\label{ambid}

We can immediately identify a number of possible effects of the spinor that persist
even if we add an extra 
fluid to the system.  Single-chirality Weyl spinors, $\phi$ or 
$\chi$, for example, are not very interesting 
in cosmology. They produce vanishing currents $A^{I}$ and $V^{I}$, with $E=B=0$. 
Such spinors therefore have no effect on the Friedmann equations, and
if there are no other matter components in the Universe, they simply 
lead to Minkowski space-time. Single-entry spinors in the Dirac representation, on the other hand, map
into solutions with $|\phi|^2=|\chi|^2$, i.e. solutions without parity
violation, where left and right spinors have the same probability. These 
``ambidextrous'' solutions do have an effect on the Friedmann equations.
They are particularly simple to integrate because they display
\be
A^0=(|\phi|^2 - |\chi|^2)=0
\ee
which at least when $U'\neq 0$ implies $B=0$ (see Eqs.~(\ref{dotb})
and (\ref{dota})). Then Eqn.~(\ref{dote}) provides the first integral:
\be\label{1intsimple}
E= \frac{M}{a^3}
\ee
which is nothing but (\ref{firstint}) when we set $B=A^0=0$.

As we see, we must distinguish between parity violation in the 
solutions to the theory, and in the theory itself (i.e. in its action), 
here only present if $\beta\neq 1/\gamma$ (cf. the last term in 
Eqn.~(\ref{4ferm})). 
Regardless of the parameters of the theory we see that solutions
with maximal parity violation (Weyl spinors) have no effect in 
cosmology, with or without torsion, and with or without parity violation in 
the actual theory. Parity invariant or ambidextrous solutions, in contrast, 
have an effect particularly simple to analyze, which we shall now do.
In between the two extremes we find intermediate solutions, harder 
to work out, which we will do numerically in Section~\ref{parviol}.

\subsection{Solutions without torsion}
If $\gamma=0$, it follows that $\xi=0$, and 
we obtain the well-known case of the torsion-free
theory. As pointed out in \cite{ArmendarizPicon:2003qk}, classical spinors are remarkable (and bypass
a number of theorems valid for scalar fields) in that 
any equation of state can be produced by appropriately designing
the potential. For example, if: 
\be 
U=\zeta\left(\bar{\Psi}\Psi\right)^{n}=\zeta E^{n}
\ee
then, since $\rho=U$ and $E\propto 1/a^3$, we have $\rho\propto 1/a^{3n}$.
Since (\ref{cons}) implies that (for a constant equation of state
$w=p/\rho$) 
$\rho\propto 1/a^{3(1+w)}$ we can read off
\be
w_0=n-1
\ee
without any further calculation. The standard results for $a(t)$ 
follow (e.g. $a\propto t^{2/(3{1+w}}$, if $w>-1$, etc).

The  $n=1$ case corresponds to the massive Dirac field, whereas $n=4/3$ 
corresponds to the G\"{u}rsey model \cite{Gursey:1956zzb}. 
In the absence of self-interactions, 
the former leads to a dust model, whereas the latter behaves like 
a radiation dominated Universe. Inflation can only be precisely obtained with 
a flat potential ($n=0$), with the problems discussed in \cite{ArmendarizPicon:2003qk}.

\subsection{Torsion driven bouncing solution}

\begin{figure}[h]
\begin{center}
\scalebox{0.7}{\includegraphics{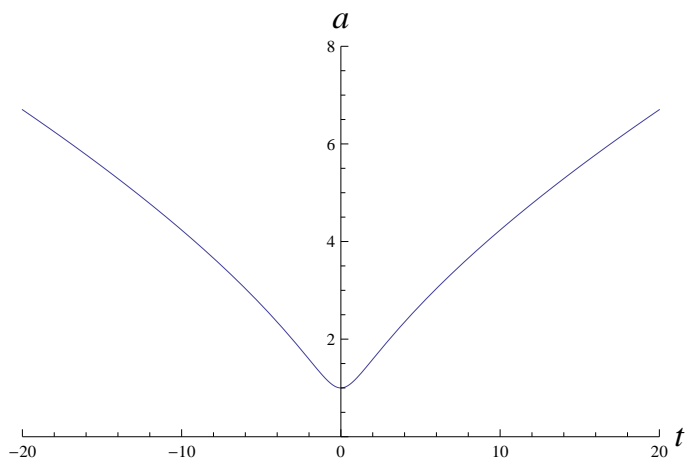}}
\caption{\label{basica}{The bounce for a typical ambidextrous solution
with $\xi<0$. In this case $U=m E$, so that the contracting phase
exhibits $a\propto (-t)^{2/3}$  and the expanding phase $a\propto t^{2/3}$.}}
\end{center}
\end{figure}

If $\gamma\neq 0$ in general we have a bouncing solution. 
For simplicity let us consider the $n=1$ mass potential, 
$U=mE$, but the results in this and the next sections
generalize to more complex potentials (although it may then be more difficult to
find analytical solutions). The case $\xi>0$ will lead
to less interesting solutions, to be reviewed in 
Section~\ref{boring}.  If $\xi<0$, the self-interactions resulting from 
torsion have the effect of dramatically reshaping the mass potential,
converting the typical mass bowl into an upside-down Mexican hat 
potential: 
\be 
W=mE-|\xi|E^2.
\ee 
Since  $\rho=W\ge 0$ (as implied by the first Friedman equation,
Eq.~(\ref{fried})) it is 
then not difficult to predict a bounce when $\rho=0$.
The only alternative would be a static universe, but
since $p\neq 0$ when $\rho=0$ this is not realized.

\begin{figure}[h]
\begin{center}
\mbox{
\subfigure{\scalebox{0.6}{\includegraphics{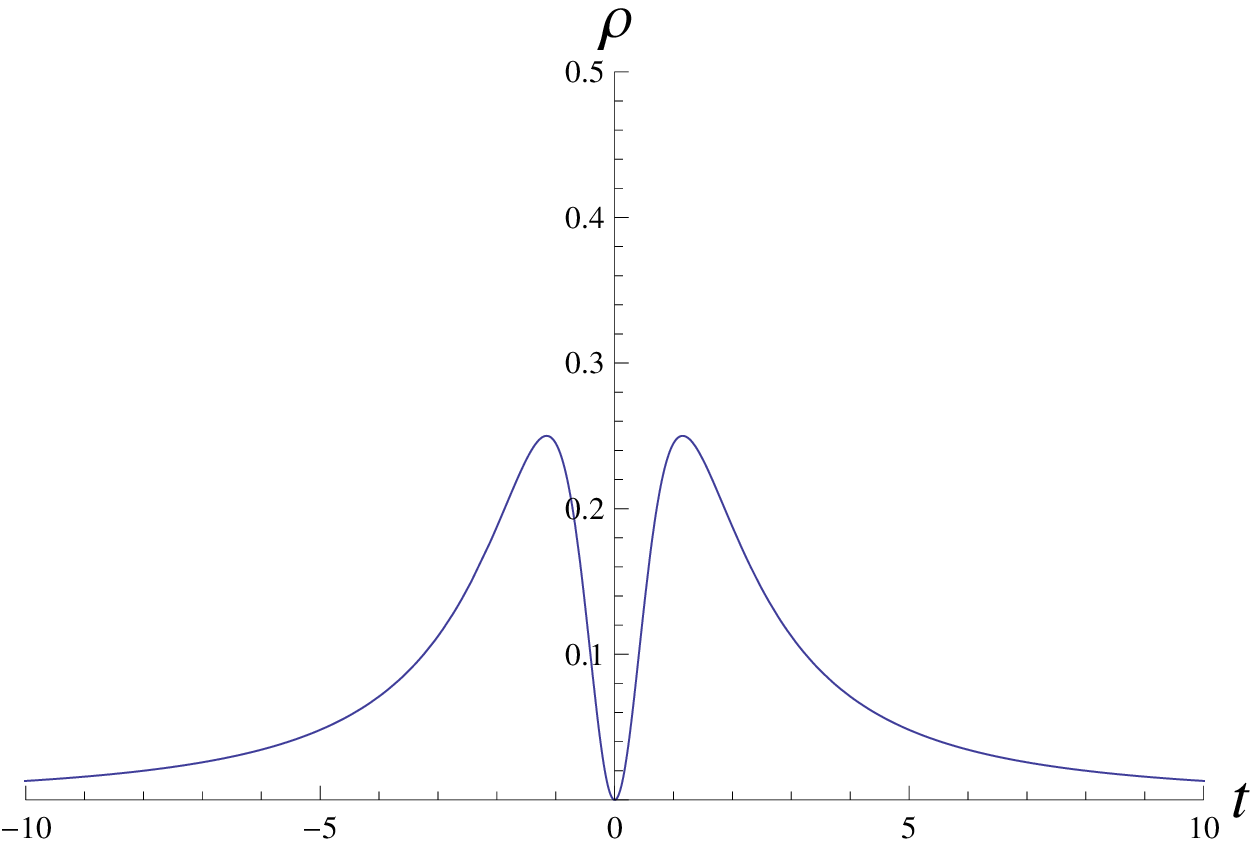}}}
\subfigure{\scalebox{0.6}{\includegraphics{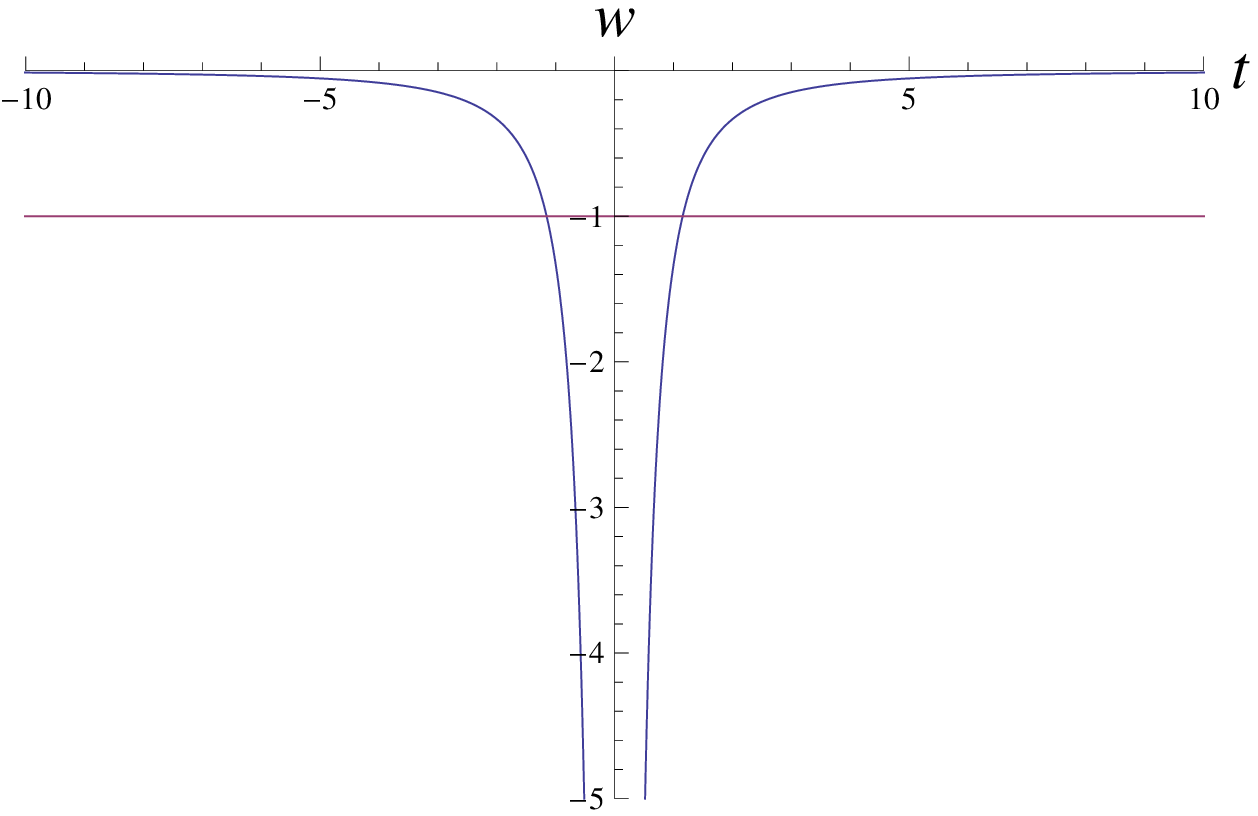}}}}
\caption{\label{basicrhow}{The values of $\rho$, displayed in units of $\rho_{0}$, and $w$ during an ambidextrous
bounce. As we can see there is a period of phantom behaviour, where the 
density decreases as the universe contracts, then increases as it expands.}} 
\end{center}
\end{figure}

We have plotted $a(t)$ in Fig.~\ref{basica}, as numerically integrated following the procedure described in Section~\ref{cosmoeqs}.  
In this particular case it is 
possible to find analytical solutions, with (\ref{fried}) and 
(\ref{1intsimple}) leading to 
\be
a(t)=\left[M\left(\frac{|\xi|}{m}+\frac{3m t^2}{4}\right)\right]^{1/3}\; ,
\ee
(and the other equations of the minimal set reading trivially $0=0$).
Asymptotically
(large $|t|$),  the Universe contracts
like  $a\propto (-t)^{2/3}$ and expands  like 
$a\propto t^{2/3}$, typical of dust. 
In between there is a phase where torsion dominates
causing a bounce. This occurs when:
\be
E=E_0=-\frac{m}{\xi}\; .
\ee
We may write 
\be 
\rho={\tilde \rho}-\frac{{\tilde\rho}^2}{\rho_0}
\ee 
where $\tilde\rho=mE$ represents the energy density before torsion 
effects are added, and
\begin{eqnarray}
\rho_{0} &=& \frac{16}{3}\frac{(\gamma^{2}+1)}{\gamma^{2}\left(\beta^{2}-\frac{2}{\gamma}\beta-1-\alpha^{2}\right)}m^{2}M_{PL}^{2}\; .
\end{eqnarray}
where we have defined the Planck mass-squared $M_{PL}^{2} \equiv 1/8\pi G$, which appears due to its contribution to the quantity $\xi$ (see equation (\ref{xi})). 
We see then that the bounce occurs when $\rho=0$ and ${\tilde \rho}=\rho_0$.
By studying further the function $\rho(\rho_0)$, 
we see that it has a maximum at 
\be
\rho_\textrm{max}=\frac{\rho_0}{4}\; . 
\ee
As the universe contracts the density increases like $1/t^2$, as usual, 
but then torsion kicks in. This maximal density is reached, and  then,
as the universe compresses further, the density {\it decreases} until 
it reaches zero and a bounce occurs at a finite $a$. After the bounce
the density at first {\it increases} with expansion, until the same
maximum $\rho_\textrm{max}$
is reached again.  After that it starts to decrease with expansion, 
eventually according to the usual $\rho\propto 1/t^2$. The density
never diverges and a Big Bang singularity is avoided. We have 
plotted this behaviour in Fig.~\ref{basicrhow}.

Increasing density with expansion (or decreasing density with contraction)
is a hallmark of phantom matter \cite{Caldwell:2003vq}. And indeed by plotting the 
equation of state $w=p/\rho$ (see Fig.~\ref{basicrhow}) we see that this
does cross $w=-1$ as the maximum density is reached. Torsion induces a phantom
period around the bounce and in fact $w$ becomes infinitely negative
at the bounce, since $p<0$ is finite while $\rho=0$. When torsion becomes
sub-dominant $w$ goes to zero, as predicted in the previous sub-section.

\section{Parity violating perturbations}\label{parviol}
\begin{figure}[h]
\begin{center}
\mbox{
\subfigure{\scalebox{0.62}{\includegraphics{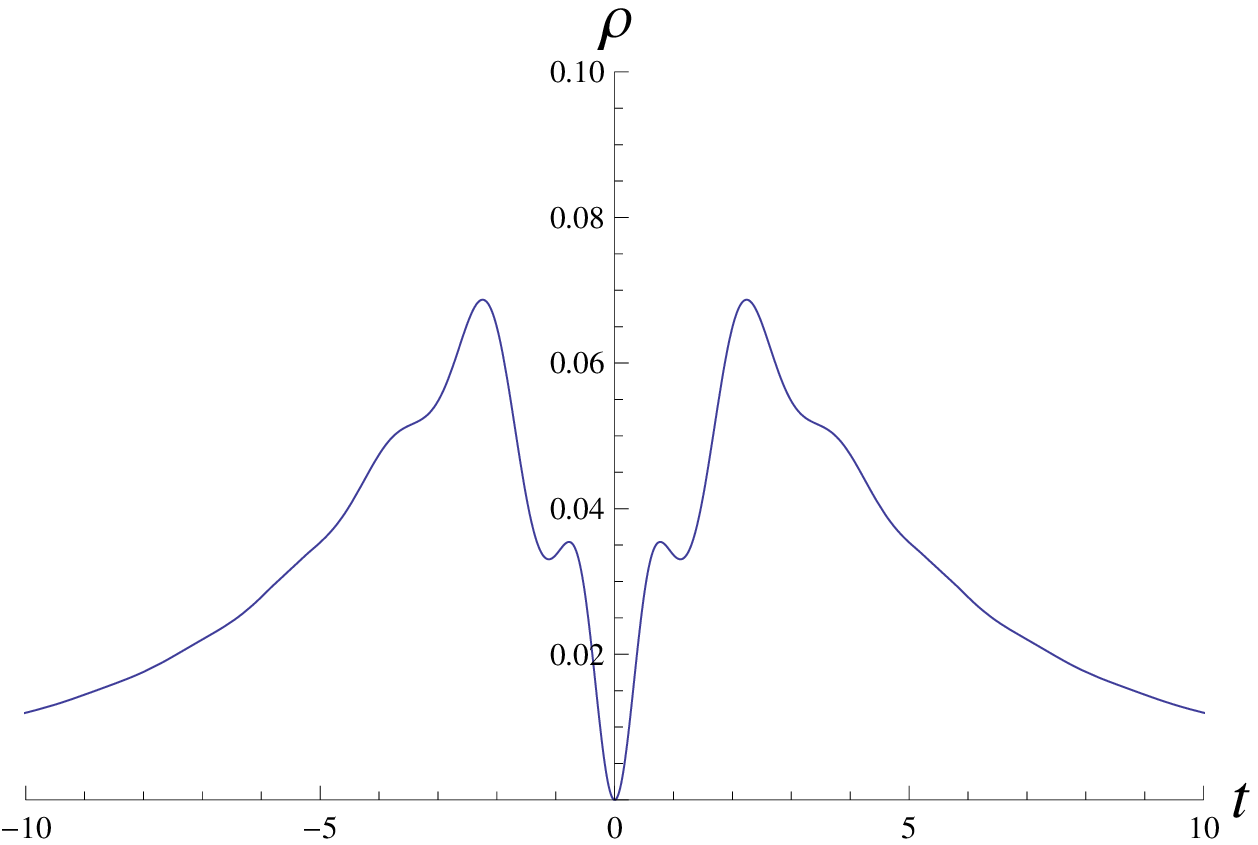}}}
\subfigure{\scalebox{0.62}{\includegraphics{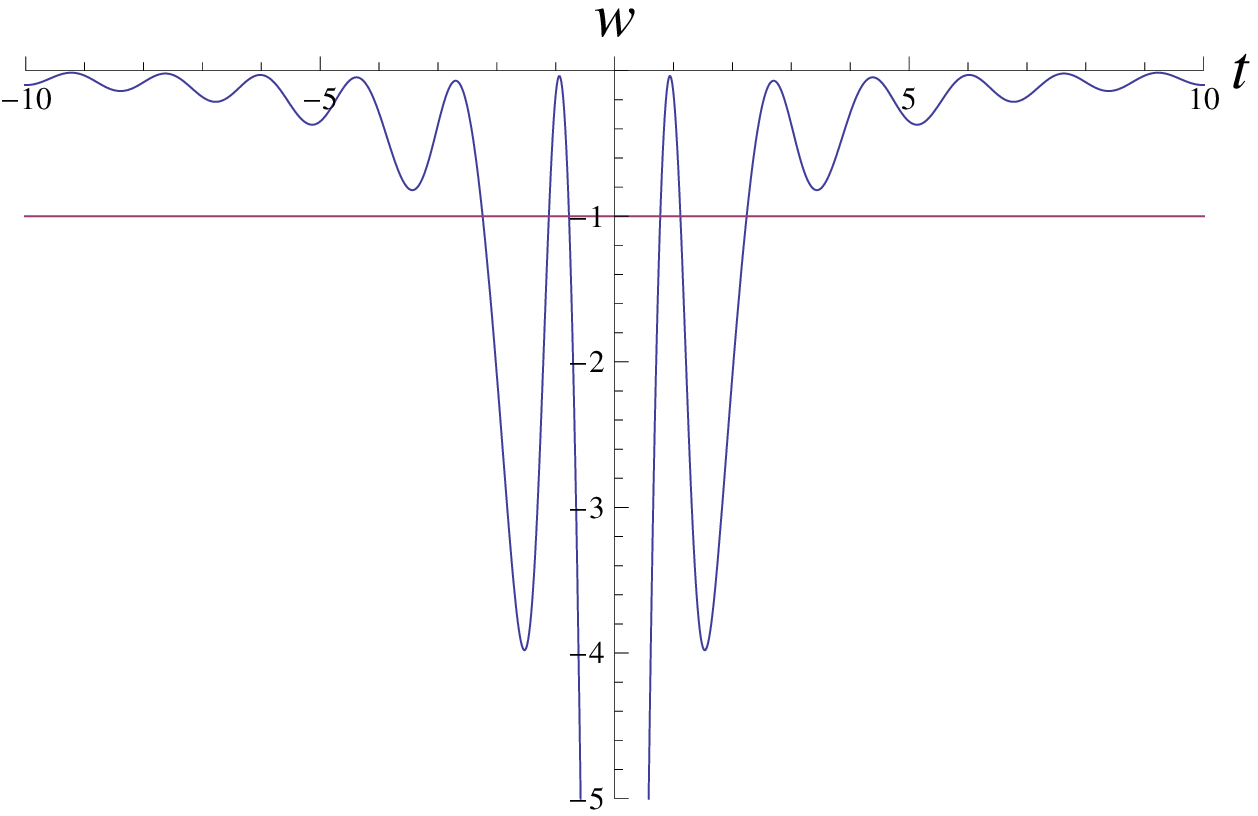}}}}
\caption{\label{Brhow}{The values of $\rho$ and $w$ with a large
$B$-type perturbation at the bounce. Even though the $a(t)$ profile
doesn't change much we find interesting oscillatory behaviour superposed
on the basic, ambidextrous picture.}} 
\end{center}
\end{figure}
The issue arises as to whether the bouncing solution presented in the 
previous Section is stable, when strict parity invariance is
broken, with $B$ and $A^0$ turned on. We will study the matter
in this Section, first turning on $B$-type perturbations at the bounce,
then $A^0$ perturbations, and then both. In order to do this we will need
to perform a numerical integration, following the procedure described in
Section~\ref{cosmoeqs}. In all cases the solution is
stable, in the sense that we find small variations in the bounce 
commensurate with the size of the perturbations induced. What is more
important, we find that even when the perturbations are very large
the overall picture does not change much in the first two cases
(pure $B$ and $A^0$ type perturbations), the solutions simply displaying
(large) oscillations around the basic bouncing solution. However,
when $B$ and $A^0$ perturbations are allowed free rein at the bounce they
introduce a very interesting qualitative novelty should they be large enough:
whilst the bounce is still present there is an asymmetry between the 
contracting and expanding phase.

\subsection{$B$-type and $A^0$-type  perturbations}
\begin{figure}[h]
\begin{center}
\mbox{
\subfigure{\scalebox{0.62}{\includegraphics{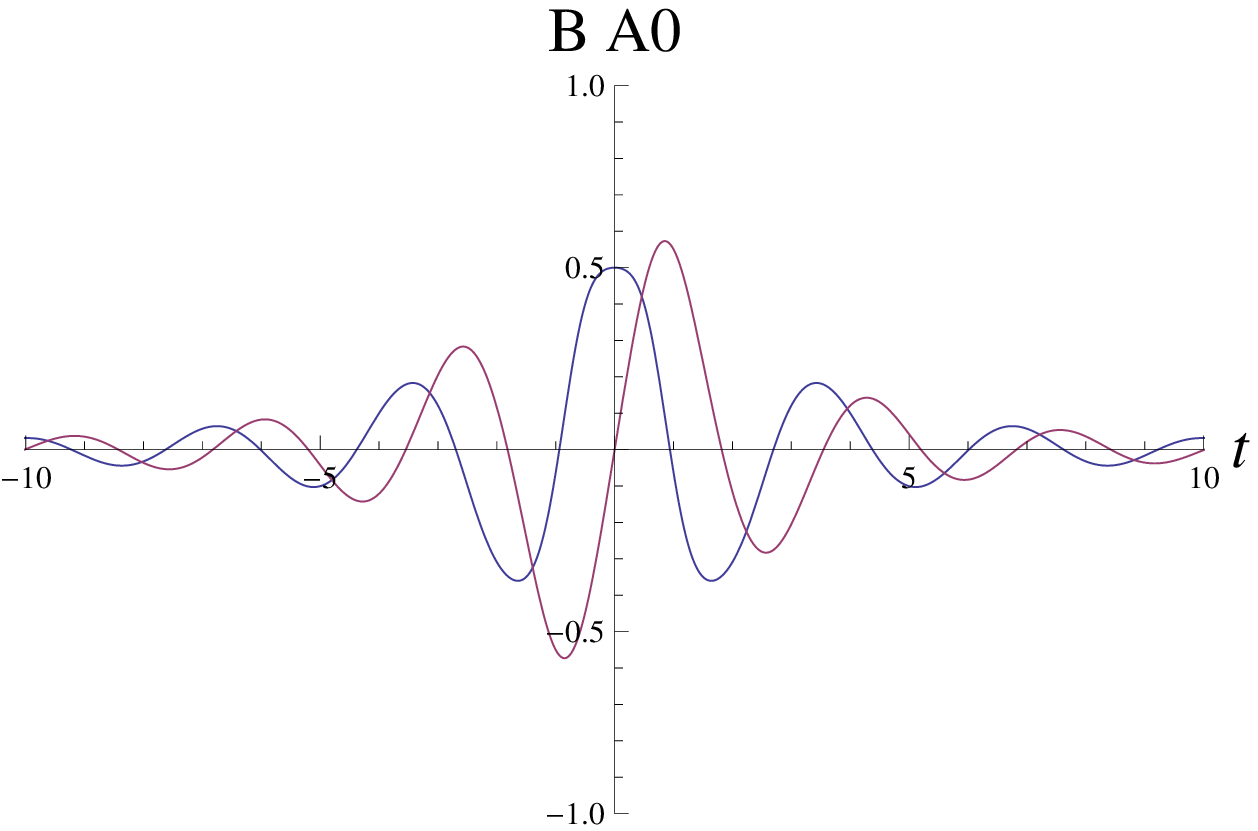}}}
\subfigure{\scalebox{0.62}{\includegraphics{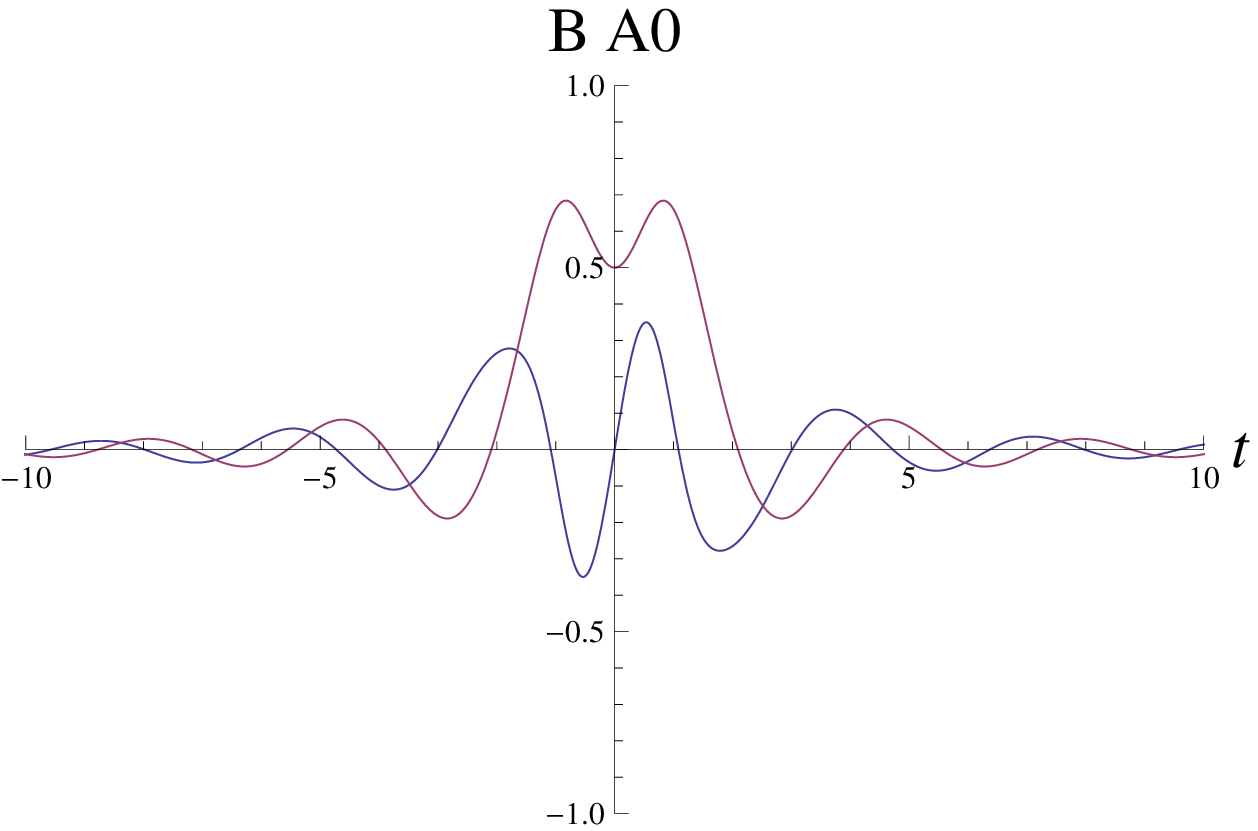}}}}
\caption{\label{BBA}{A plot of $B$ (blue) and $A^0$ (red) for $B$-type and $A$-type
large perturbations. }} 
\end{center}
\end{figure}

As just stated, we do not see any instability in the bouncing solution,
and small additions of $B$ and $A^0$ lead to perturbations of the same
order. What is interesting is that even if the perturbations
are very large (order 1 and higher) the overall picture does not
qualitatively change, as long as one of $B$ and $A^0$ are zero at the bounce.
If $B=0$ at the bounce we call this an $A^0$-type perturbation and 
vice-versa.
In Fig.~\ref{Brhow} we plot the effect of a large $B$-type perturbation 
on $\rho$ and $w$. The overall picture is essentially
the same,  with oscillatory behaviour superposed
on the ambidextrous picture. This is due to the oscillatory nature
of the variables $B$ and $A^0$ (see Fig.~\ref{BBA}, top). Obviously
the value for the maximal density is now different, and most
easily determined numerically, but the basic picture remains. 
The bounce occurs when 
\be
E=\frac{-m\pm {\sqrt{m^2-4\xi^2 B^2}}}{2\xi}\; .
\ee
The picture is similar for an $A^0$-type
perturbation. In Fig.~\ref{BBA} we plot the behaviour of $B$ and $A^0$ 
for $B$-type and $A$-type large perturbations, for comparison.
The two variables are generally out of phase, and the modes considered
here correspond to one of them having a node at the bounce.
We don't plot the $a(t)$ profile because this is basically 
indistinguishable form the unperturbed case described in Fig.~\ref{basica}. 


\begin{figure}[h]
\begin{center}
\scalebox{0.7}{\includegraphics{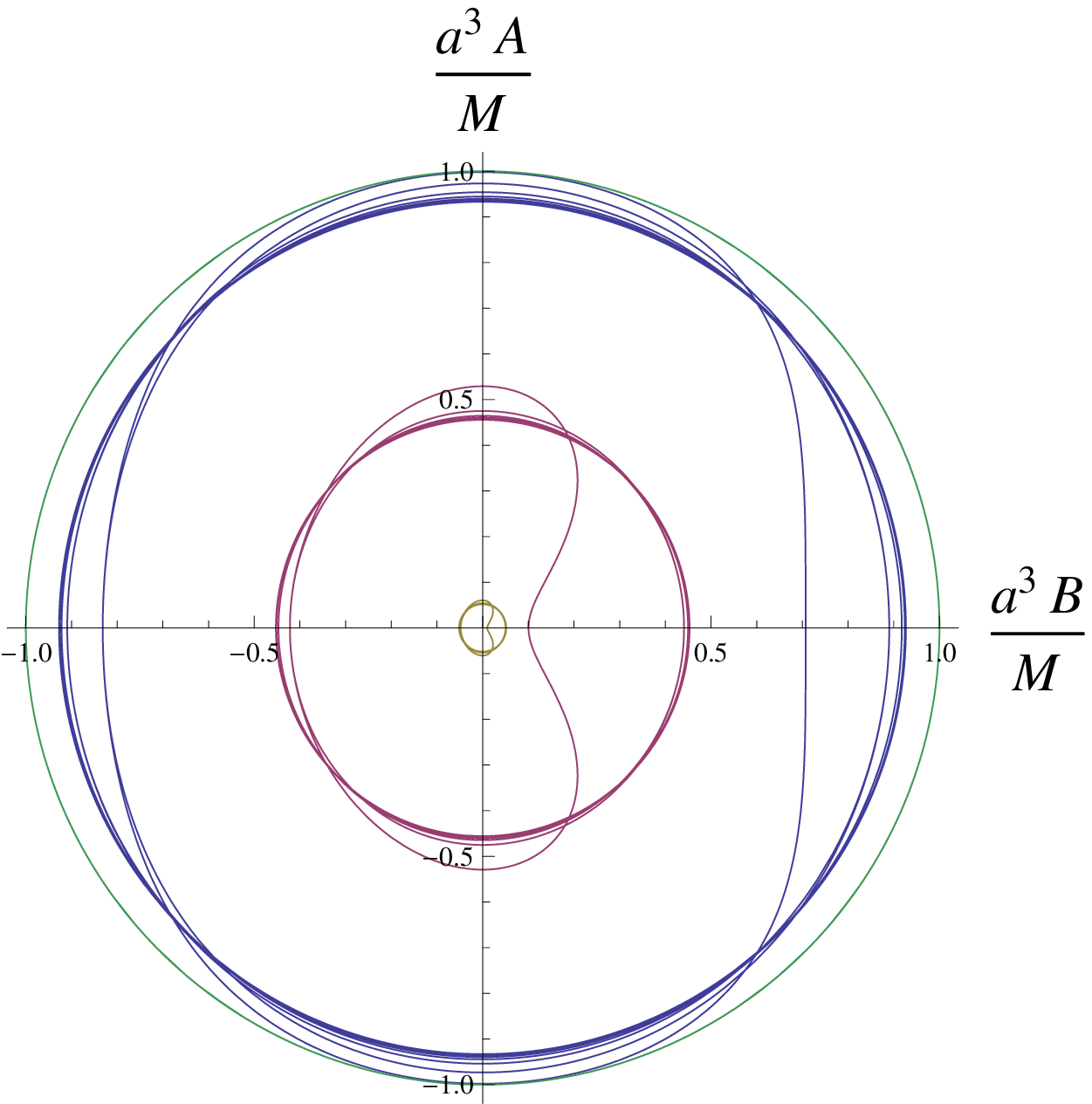}}
\caption{\label{Bpplot}{The $\{x,y\}$ trajectories of $B$-type
perturbations (trajectories constrained to remain inside
the unit circle, depicted). The outer trajectory corresponds to the 
large perturbations used in the previous plots. The center represents the 
exactly ambidextrous solution presented in the last Section. The intermediate
trajectories correspond to smaller and smaller perturbations in the 
initial conditions. As we see the ambidextrous solution is stable, but
furthermore, even very large perturbations arrange themselves as 
oscillations around this solution. }}
\end{center}
\end{figure}

\begin{figure}[h]
\begin{center}
\scalebox{0.7}{\includegraphics{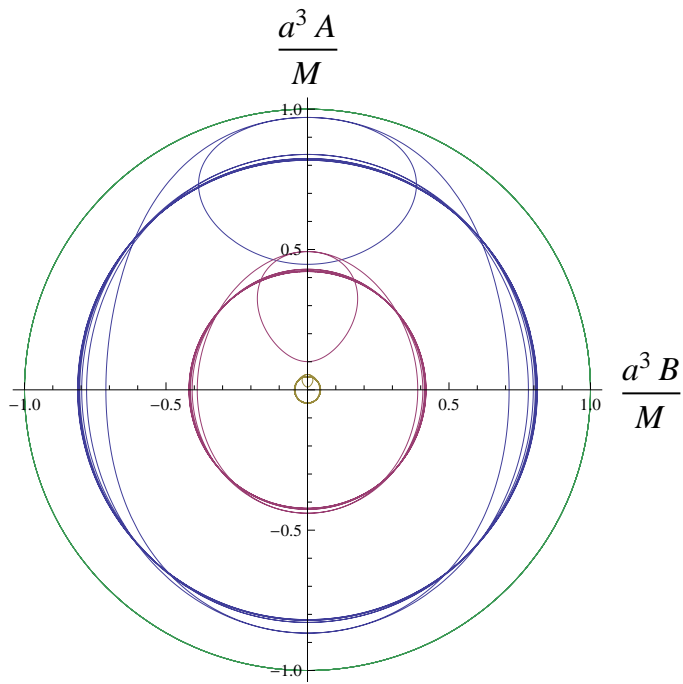}}
\caption{\label{Applot}{The equivalent of Figure~\ref{Bpplot} for an 
$A$-type perturbation. }}
\end{center}
\end{figure}

Another way to represent these results is to map the dynamics onto the plane
spanned by
\bea
x&=&\frac{a^3 B}{M}\\
y&=&\frac{a^3 A^0}{M}\; .
\eea
Given Eq.~(\ref{firstint}), 
the system is constrained to remain inside the unit circle, with the
distance to the boundary providing a measure of $a^3 E$. 
In this plot, the origin corresponds to the exactly ambidextrous solution 
presented presented in the last Section. The $\{x,y\}$ trajectories 
away from the origin represent the parity violating perturbations. 
In Fig.~\ref{Bpplot} and~\ref{Applot} we show these trajectories
for $B$-type and $A$-type perturbations, respectively.
The outer trajectories correspond to the rather large perturbations used 
in the previous plots. The inner trajectories correspond to 
smaller and smaller perturbations in the  initial conditions. 
As we see the ambidextrous solution is stable; but more interestingly
even very large perturbations arrange themselves as oscillations
around this solution, which therefore seems a generic feature. 


\subsection{Generic perturbations}
\begin{figure}[h]
\begin{center}
\mbox{
\subfigure{\scalebox{0.65}{\includegraphics{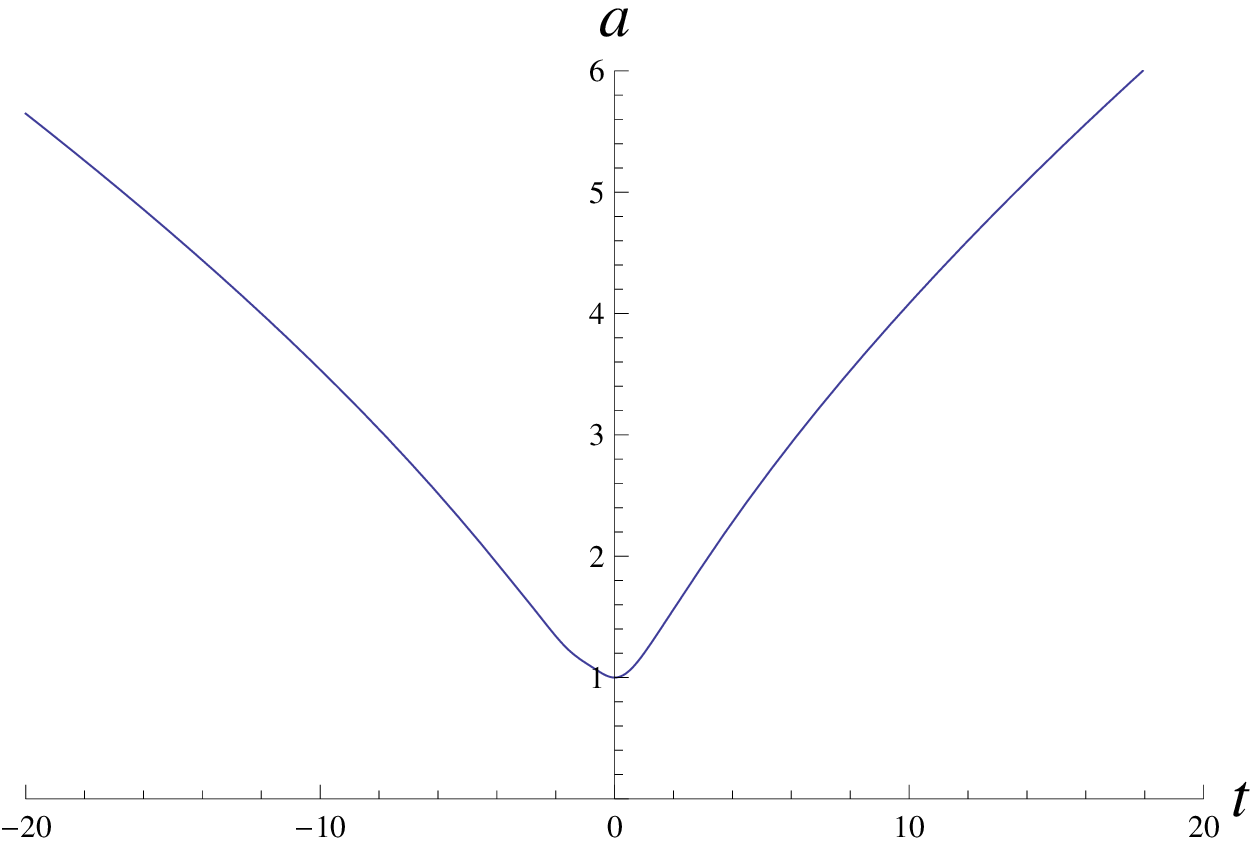}}}
\subfigure{\scalebox{0.65}{\includegraphics{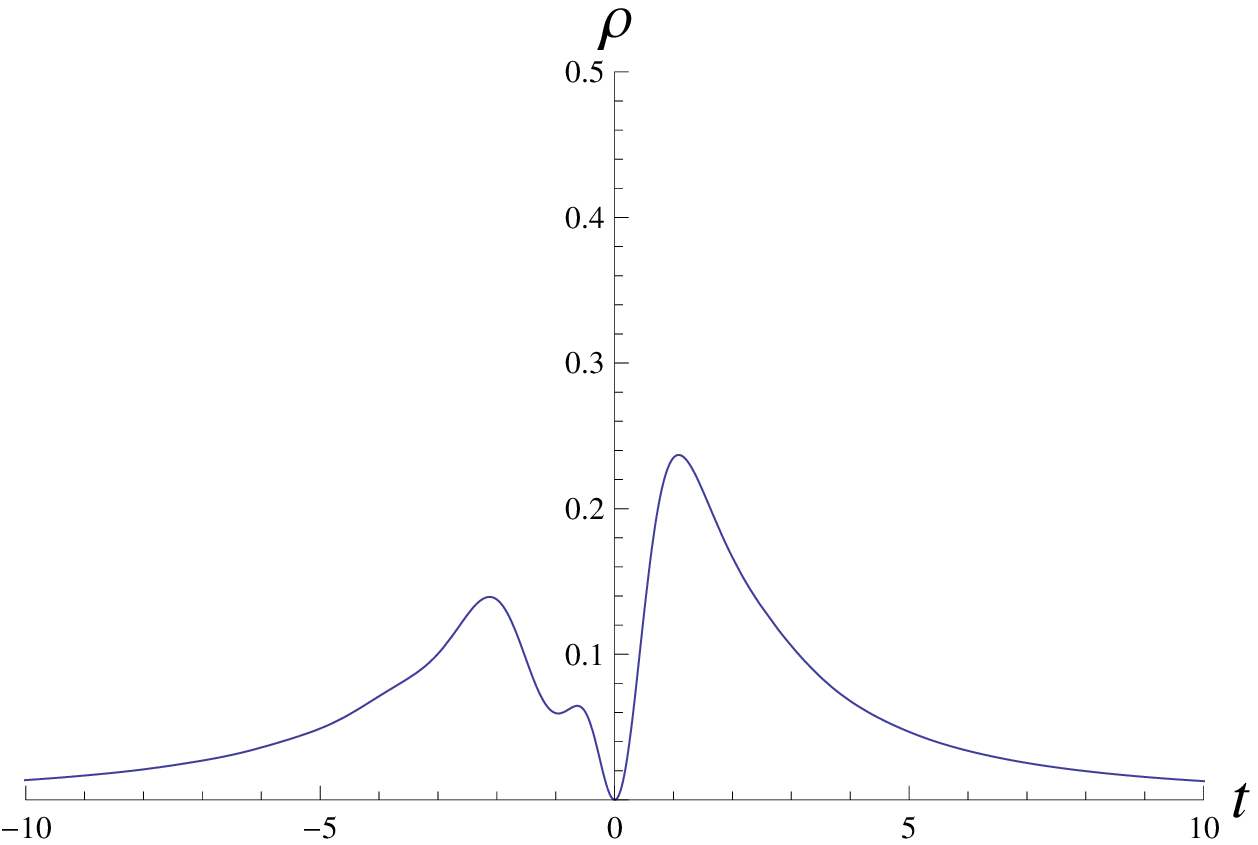}}}}
\caption{\label{BAarho}{For general large perturbations (where neither 
$B$ nor $A^0$ have a node at the bounce) we observe an asymmetry between
the contracting and expanding phase, with the maximal density different
in the two phases.}}
\end{center}
\end{figure}

\begin{figure}[h!]
\begin{center}
\scalebox{0.7}{\includegraphics{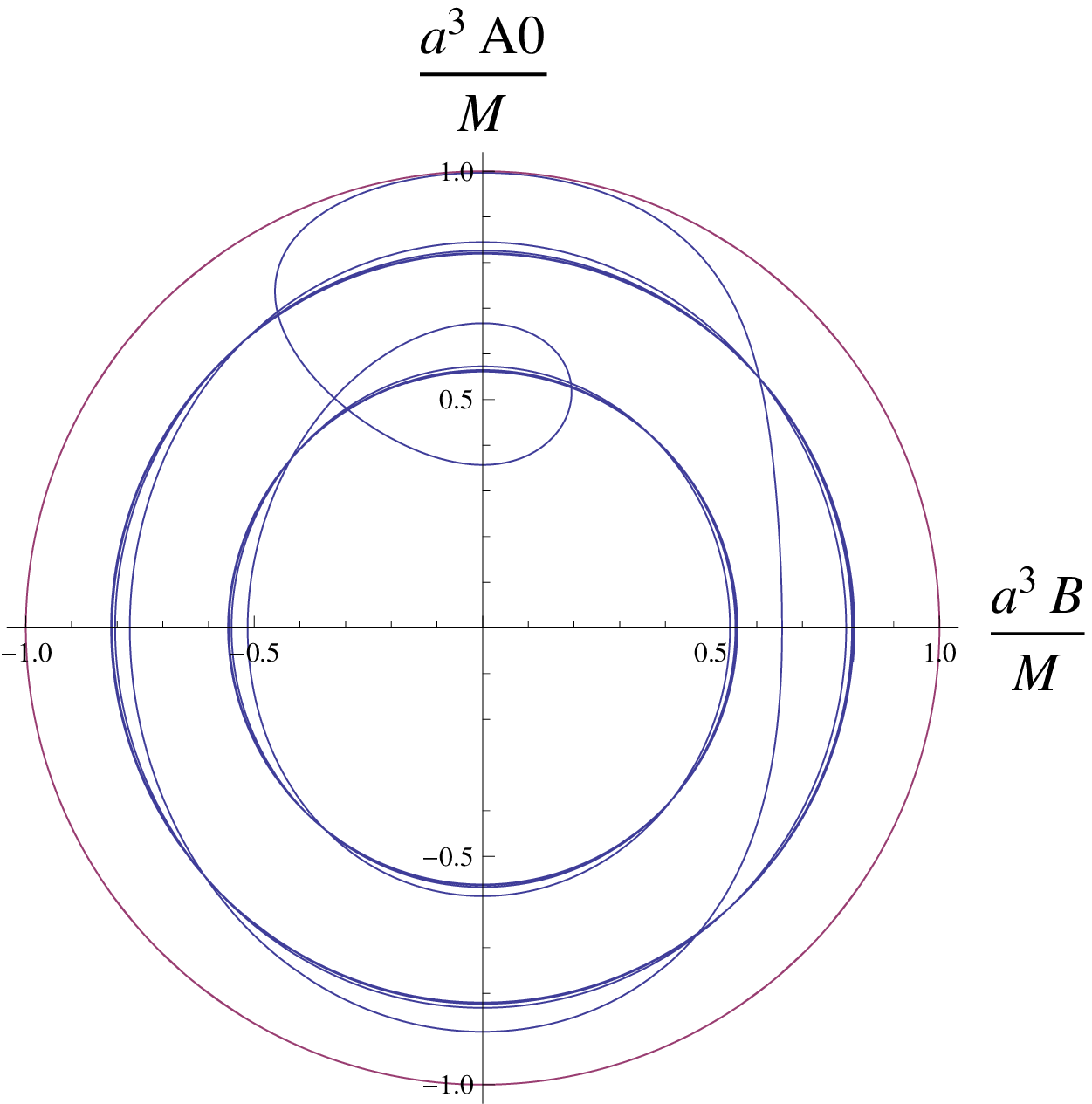}}
\caption{\label{BApplot}{The $\{x,y\}$ diagram for the evolution of the 
system under a generic large parity violating perturbation. We see that 
the two circles, corresponding to the period when torsion has died down,
differ for the contracting and expanding phase. The evolution during the 
bounce is also more complex and asymmetric.}}
\end{center}
\end{figure}

The only qualitative novelty appears if we mix the two types of perturbations,
and let their amplitude be very large.
If we set up the perturbations away from the bounce in general it does not
happen that one of $B$ or $A^0$ vanish at the bounce. Then, 
it still happens that the ambidextrous solution is stable against
small perturbations, and that even very large perturbations
consist of oscillations around this solution.  However a novelty appears: 
there appears an asymmetry between the contracting and expanding phase, as we
illustrate in Fig.~\ref{BAarho}. For example, the maximal energy density 
reached is different in the two phases, as is the  
defining ``constant''  $a^3 \rho$
associated with the two dust Universes. Depending on the initial conditions
chosen, this may be smaller or larger in the expanding phase. 
We also depict the evolution in an $\{x,y\}$ plot  in Fig.~\ref{BApplot}. 
The evolution is still an oscillation around the ambidextrous
solution (origin) but now the amplitude on either side of the bounce is 
different.

\subsection{Other solutions}\label{boring}
There are a number of other solutions which are not particularly interesting,
but which we list here for completeness. 

Should $\xi>0$, then there isn't a bounce and the singularity is 
not avoided. Indeed as we go back in time, the singularity is reached faster
because the torsion adds to the pressure. The dust phase described
above is then preceded by a period of kination (to borrow terminology
from scalar field cosmologies), i.e. a period with $\rho\propto
1/a^6$ and $a\propto t^{1/3}$. This is true for ambidextrous solutions,
but does not qualitatively change for more general solutions. 
If the theory is massless there is only a period of kination,
and this happens for the more general case with non-vanishing $B$.
In that case $B\propto 1/a^3$ and it does not affect this conclusion. 
It could also be that $U=U_0$ is a constant, in which case the period
of kination with $E$ and $B$ decaying like $1/a^3$ is followed by a 
de Sitter phase. 
None of this is very surprising or interesting. 
The strength of this
paper is in the $\xi<0$ solutions.

\section{Conclusions}\label{conclusions} 

We conclude with an appraisal of what we have discovered and what 
remains to be done. We found that the overall picture of cosmologies
driven by spin hinges on the sign of one combination of parameters, $\xi$
as defined in Eq.~(\ref{xi}). 
If $\xi>0$, as is the case of Einstein-Cartan theory with minimal coupling, nothing dramatically 
new happens. For example, for a mass potential (or any potential leading 
to $w<-1/3$) the only novelty is that an early period of 
kination precipitates the onset of a singularity as we go back in time. 
But if $\xi<0$ the singularity is avoided and even for a simple 
mass potential a bounce is generic, even against very large
perturbations of the basic solution. This double picture is closely related 
to the  attractive/repulsive nature of the 4-fermion interaction. As is well
known, in the Einstein-Cartan theory with minimal coupling this interaction is attractive
(at least for classical spinors), with the result that
singularities are easier to 
form than in the torsion-free theory~\cite{Kerlick:1975tr}. 
By identifying a non-minimal coupling leading to $\xi<0$ we have  
reversed the situation, rendering the spinor self-interaction
repulsive, and opening up the doors to
singularity avoidance. It is curious to note
that such scenarios are closely related to the presence of 
parity-violating terms in the action.

Several questions can be raised. We presented an extensive
set of solutions, displaying simple and complicated bounces which may be 
symmetric or not. But have we found all possible solutions? 
It is tempting to speculate on the existence of 
a de Sitter/inflationary fixed point. Indeed the 
effects of torsion when $\xi<0$ and $B\neq 0$ are qualitatively similar 
to those of the C-field in steady-state cosmology: a sea of
negative energy (here represented by torsion) which interacts with a 
positive energy component (when $B\neq 0$ and $A^0\neq 0$). 
Such situations often lead to inflation, i.e. a sustained period of 
accelerated expansion. We have been
unable to find such a solution. The reason is probably that in our case
the same field has both a positive and a negative energy component, so 
the situation is not quite the same. Nonetheless we 
defer to a future paper a more complete analysis, based on 
phase space portraits of autonomous dynamical
systems. We stress that there is accelerated expansion at the bounce
($\ddot a>0$) but this is not inflation, which is a {\it sustained} period 
of accelerated expansion (this seems to have been missed 
in~\cite{Ribas:2009yg}). The possibility cannot be dismissed, however, 
that a long inflationary transient is present somewhere in the phase space.

Is the bounce we have found stable against anisotropy domination?
This is a valid concern because even for small perturbations
the shear tensor $\sigma_{ij}$ contributes to the Friedmann equation
as $\sigma^2=\sigma_{ij}\sigma^{ij}/2$, and $\sigma\propto
1/a^3$. Therefore, unless the equation of state is super-stiff ($w>1$)
the anisotropy tends to dominate during the contraction, leading to a 
mixmaster phase or even a singularity, rather than a smooth bounce. 
In our case we are in a borderline situation, since {\it if torsion and
shear do not interact} then they both scale as $1/a^6$. Therefore the solution
we have presented is certainly stable against small, but not large
shear perturbations. However the situation may be more subtle, and
we defer further analysis to a future publication. As we noted before,
a spinor field already is anisotropic; however this need not be reflected
in the metric (at least if $K=0$).

Even ignoring the issue of anisotropy, the obvious next step is 
to work out the fluctuations in this type of model, finding the amplitude,
spectral index and tensor/scalar ratio as a function of the free parameters
of the model. This has been examined in the past for spinor-driven 
cosmologies in the context of 
inflation~\cite{ArmendarizPicon:2003qk,Watanabe:2009nc,Cai:2011tc} 
(which may be achieved by making $U$ very flat). As 
the work of~\cite{ArmendarizPicon:2003qk} shows, spinors and scalar
fields are very different in this respect: for the same background
kinematics one gets a spectral index $n_S=4$ for the former where for
the latter $n_S =1$ is found. For this reason the status of our model
regarding fluctuations is far from obvious. 
It is interesting to point out that for 
scalar fields a $w=0$ (dust-like) bounce does have a 
scale-invariant mode~\cite{Wands:1998yp,Finelli:2001sr}. 
However these models suffer from fine-tuning problems related to the
fact that the spectrum of the curvature and potential fluctuations is not the 
same. In future work we hope to examine how this might
change if the bounce were to be driven by a spinor field. In this work we will also consider a more realistic scenario, where to the spinor field a radiation component is added. This does not qualitatively change any of the conclusions in this paper regarding the bounce. But it will affect its details, and in fact it will be essential for a proper description of fluctuations in these models.

The major concern remains as to what might be the physical basis for 
spinor models
(but this criticism could be levelled at most early Universe models,
including those based on scalar fields). It obviously would be more 
conservative to take the ``spin-fluid'' approach, such as that pioneered
by Weyssenhoff (see for 
example~\cite{Gasperini:1986mv,Gasperini:1998eb,Boehmer:2006gd,Poplawski:2010kb,Poplawski:2011jz,Poplawski:2012ab} for some very interesting cosmological work based on this
approach\footnote{Note though that there appears to be an essential friction between the Weyssenhoff spin fluid and any underlying theory based on minimally coupled fermions. The reason for this is that the object $C_{IJK}$ does not possess the same symmetries in the two cases \cite{Bauerle:1983ai}}), or by considering the cosmological effect of thermalized fermionic matter \cite{Dolan:2009ni,Diakonov:2011fs}. However, it may be argued that it is also legitimate to consider a spinor field as envisaged here (see Appendix 
of~\cite{ArmendarizPicon:2003qk}, for example). In addition we note
that the theory can be phrased wholly in terms of the bilinear invariants
(such as the ODEs presented in Section~\ref{cosmoeqs}). With this 
remark in mind in future work we hope to refine the argument 
in~\cite{ArmendarizPicon:2003qk}. Notice that the sign issues presented
in Section~\ref{signals} change dramatically depending on whether 
the underlying field can be considered to be classical (for example $\Psi^\dagger\Psi$
is positive definite classically but $\left<\Psi^\dagger\Psi\right>$ of course need not be so in the quantum theory). It is interesting to note that perhaps the most in-depth analysis to date of classical spinors at the level of the cosmological background and cosmological perturbations  has been in the context of non-standard/dark spinors \cite{Boehmer:2007ut,Boehmer:2009aw,Boehmer:2010ma}.

We close by pointing out that the model presented here
has equations very similar to those found in loop quantum cosmology
and the brane-world scenario, but only when $B=0$. If $B\neq 0$ we seem
to be generalizing the dynamics in those models.

\section{Acknowledgments} 
\noindent We would like to thank Stephon Alexander, John Barrow, 
Laurent Freidel, Friedrich Hehl, Antonino Marcian\`o, Ugo Moschella, 
and Hans Westman for helpful discussions and an anonymous referee for very useful suggestions. This work was funded by STFC through a consolidated grant.

\appendix
\section{Conventions}
\label{conventions}

In this section we describe the conventions used in this paper for various quantities. Often this information is omitted
in papers about the role of torsion in cosmology, making comparison of results difficult.

We use distinct symbols $\epsilon_{IJKL}$ and $\varepsilon^{\mu\nu\sigma\delta}$. The object $\epsilon_{IJKL}$ is a spacetime scalar antisymmetric in all indices and with $\epsilon_{0123}=1$; the object is invariant under local $SO(1,3)$ transformations. 
The object $\varepsilon^{\mu\nu\sigma\delta}$ is a spacetime \emph{density} antisymmetric in all indices and with $\varepsilon^{0123}=1$; the object is numerically invariant under local coordinate transformations. The determinant $e$ of the co-tetrad $e^{I}=e^{I}_{\mu}dx^{\mu}$ is subsequently defined as follows:

\begin{eqnarray}
e &=& \frac{1}{4!}\epsilon_{IJKL}\varepsilon^{\mu\nu\delta\sigma}e_{\mu}^{I}e_{\nu}^{J}e_{\delta}^{K}e_{\sigma}^{L}
\end{eqnarray}
and hence

\begin{eqnarray}
\varepsilon^{\mu\nu\delta\sigma}e_{\mu}^{I}e_{\nu}^{J}e_{\delta}^{K}e_{\sigma}^{L}= e\varepsilon^{IJKL} = -e\epsilon^{IJKL}
\end{eqnarray}

Furthermore we use the convention that $\eta_{IJ}=\mathrm{diag}(-1,1,1,1)$ which implies that the spacetime metric $g_{\mu\nu}=\eta_{IJ}e^{I}_{\mu}e^{J}_{\nu}$ has mostly positive signature. In using spinors in this paper we will use the 
Weyl/chiral representation i.e.

\begin{eqnarray}
\Psi= \left(\begin{array}{c}
\phi_{a}\\ 
\chi^{a'}\end{array}\right)
\end{eqnarray}
where $a$ and $a'$ are indices of the left and right handed representation of $SL(2,C)$ respectively \cite{Palmer:2011bt}.
We choose the following convention for the gamma matrices $\gamma^{I}$:

\begin{eqnarray}
\gamma^I  & = &
 \left(
\begin{array}{cc}
0 & \left(\sigma^I\right)_{aa'} \\
\left(\bar{\sigma}^{I}\right)^{a'a} & 0
\end{array}     \right) ,
 \gamma^5= \left(
\begin{array}{cc}
-1 &   0\\
0 &   1
\end{array}     \right)  , \nonumber 
\end{eqnarray}

\begin{eqnarray}
\bar{\sigma}^{I} & = &   ({\bf 1} ,-\sigma^{i}) , \nonumber \\
\sigma^{I} &=& ({\bf 1},\sigma^{i})\nn
\end{eqnarray}
where the $\sigma^{i}$ are the Pauli sigma matrices. The $spin(1,3)$ generators ${\cal J}^{IJ}$ take the following form:

\begin{eqnarray}
{\cal J}^{IJ} &=& -\frac{i}{4}\left[\gamma^{I},\gamma^{J}\right]
\end{eqnarray}
We now detail our curvature conventions. Our basic object representing curvature is the two-form $R^{IJ}$:

\begin{eqnarray}
R^{IJ}  &=& d\omega^{IJ} + \omega^{I}_{\ph{I}K} \omega^{KJ}
\end{eqnarray}
We can define the `orthonormal components' $R^{IJ}_{\ph{IJ}KL}$ via the following relation:

\begin{eqnarray}
R^{IJ} = \frac{1}{2}R^{IJ}_{\ph{IJ}\mu\nu}dx^{\mu}\w dx^{\nu} &=& \frac{1}{2}R^{IJ}_{\ph{IJ}KL} e^{K}_{\mu}e^{L}_{\nu} dx^{\mu} \w dx^{\nu}\\
&=& \frac{1}{2}R^{IJ}_{\ph{IJ}KL}e^{K} e^{L}
\end{eqnarray}
We define the Ricci tensor ${\cal R}_{\mu\nu}$ as follows:

\begin{eqnarray}
{\cal R}_{\mu\nu} &=& e_{I\mu}e^{J}_{\nu}R^{IK}_{\ph{IK}JK}
\end{eqnarray} 
Consequently the Ricci scalar ${\cal R} \equiv R^{\mu}_{\mu}$ is defined as:

\begin{eqnarray}
{\cal R} &=& R^{JK}_{\ph{JK}JK}
\end{eqnarray}
and we define the Einstein tensor $G_{\mu\nu}$ as:

\begin{eqnarray}
G_{\mu\nu} &\equiv & {\cal R}_{\mu\nu}-\frac{1}{2} {\cal R}g_{\mu\nu}
\end{eqnarray}

\section{Actions In Standard Notation} 
\label{standard}

We now use the results of the previous section to write our actions in conventional notation. This will make it easier to make contact with antecedent results in the literature. We first begin with the gravitational action:

\begin{eqnarray*}
S_{G} &=& \kappa \int \left(\epsilon_{IJKL} +\frac{2}{\gamma}\eta_{IK}\eta_{JL}\right)e^{I} e^{J}  R^{KL}\\
   &=& \frac{\kappa}{2} \int \left(\epsilon_{IJKL} + \frac{2}{\gamma}\eta_{IK}\eta_{JL}\right)R^{KL}_{\ph{JK}MN} e^{I} e^{J} e^{M}  e^{N} \\
  &=& \frac{\kappa}{2} \int \left(\epsilon_{IJKL} +\frac{2}{\gamma}\eta_{IK}\eta_{JL}\right)R^{KL}_{\ph{JK}MN} e^{I}_{\mu} e^{J}_{\nu} e^{M}_{\delta} e^{N}_{\sigma} dx^{\mu}\w dx^{\nu} \w dx^{\delta} \w dx^{\sigma}\\
  &=& \frac{\kappa}{2} \int \left(\epsilon_{IJKL} +\frac{2}{\gamma}\eta_{IK}\eta_{JL}\right)R^{KL}_{\ph{JK}MN} e^{I}_{\mu} e^{J}_{\nu} e^{M}_{\delta} e^{N}_{\sigma}\varepsilon^{\mu\nu\delta\sigma} d^{4}x\\
  &=& \frac{\kappa}{2} \int \left(\epsilon_{IJKL} + \frac{2}{\gamma}\eta_{IK}\eta_{JL}\right)R^{KL}_{\ph{JK}MN}\varepsilon^{IJMN} e d^{4}x \\
  &=& \kappa \int \left(\left(\delta_{K}^{M}\delta_{L}^{N}-\delta_{L}^{M}\delta_{K}^{N}\right)
  +\frac{1}{\gamma}\varepsilon_{KL}^{\ph{KL}MN}\right)R^{KL}_{\ph{JK}MN} ed^{4}x\\
  &=& \kappa \int \left( 2R^{MN}_{\ph{MN}MN}+\frac{1}{\gamma}\varepsilon_{KL}^{\ph{KL}MN}R^{KL}_{\ph{JK}MN} \right)ed^{4}x\\
  &=& \kappa \int \left(2{\cal R}+\frac{1}{\gamma}\varepsilon_{KL}^{\ph{KL}MN}R^{KL}_{\ph{JK}MN} \right)ed^{4}x
\end{eqnarray*}
Clearly then we have $\kappa= 1/32\pi G$. Next we write the spinor action $S_{\Psi}$ in a more familiar form. We first consider the the action in the limit of zero torsion $T^{I}=0$:

\begin{eqnarray*}
S_{\Psi,T^{I}=0} &=& \frac{1}{3!}\int \frac{i}{2}\epsilon_{IJKL} e^{I} e^{J} e^{K}  \bar{\Psi}\gamma^{L}D\Psi-\frac{U}{4}\epsilon_{IJKL} e^{I} e^{J}  e^{K} e^{L}-h.c\\
&=& \frac{1}{3!}\int\left( \frac{i}{2}\epsilon_{IJKL} e^{I}_{\mu}e^{J}_{\nu}e^{K}_{\delta}\bar{\Psi}\gamma^{L}D_{\sigma}\Psi-\frac{U}{4}\epsilon_{IJKL}e^{I}_{\mu} e^{J}_{\nu} e^{K}_{\delta} e^{L}_{\sigma}\right)\varepsilon^{\mu\nu\delta\sigma} d^{4}x-h.c\\
&=& \frac{1}{3!}\int\left( \frac{i}{2}\epsilon_{IJKL} e^{I}_{\mu}e^{J}_{\nu}e^{K}_{\delta}e^{M}_{\sigma}e_{M}^{\alpha}\bar{\Psi}\gamma^{L}D_{\alpha}\Psi-\frac{U}{4}\epsilon_{IJKL} e^{I}_{\mu} e^{J}_{\nu} e^{K}_{\delta} e^{L}_{\sigma}\right)\varepsilon^{\mu\nu\delta\sigma} d^{4}x-h.c\\
&=& \frac{1}{3!}\int\left( \frac{i}{2}\epsilon_{IJKL} \varepsilon^{IJKM}e_{M}^{\alpha}\bar{\Psi}\gamma^{L}D_{\alpha}\Psi-\frac{U}{4}\epsilon_{IJKL} \varepsilon^{IJKL}\right)e d^{4}x-h.c\\ 
&=& \int \left(\frac{i}{2}\delta^{M}_{L}e_{M}^{\alpha}\bar{\Psi}\gamma^{L}D_{\alpha}\Psi-U \right) ed^{4}x-h.c \label{fourfact}\\
&=& \int\left(\frac{i}{2}e_{L}^{\mu}\bar{\Psi}\gamma^{L}D_{\mu}\Psi- U \right) ed^{4}x-h.c
\end{eqnarray*}
Finally we turn to the non-minimal coupling terms of the left hand side of equation (\ref{act1}). In terms of our Dirac spinor $\Psi$ (and recalling the redefinition of $\chi$ following (\ref{act1})) this action, $S_{nm}$ can be written:

\begin{eqnarray*}
S_{nm} &=& \frac{1}{12}\int \epsilon_{IJKL}e^{I} e^{J} e^{K} \left(\alpha\bar{\Psi}\gamma^{L}D\Psi+\beta\bar{\Psi}\gamma^{5}\gamma^{L}D\Psi\right)+h.c \\
 &=& \frac{1}{2}\int e^{\mu}_{L}\left(\alpha\bar{\Psi}\gamma^{L}D_{\mu}\Psi+\beta\bar{\Psi}\gamma^{5}\gamma^{L}D_{\mu}\Psi\right)e d^{4}x + h.c
\end{eqnarray*}

\bibliographystyle{hunsrt}
\bibliography{references}

\end{document}